\documentclass[twocolumn,amsmath,amssymb,aps,pra,floatfix]{revtex4-1}

\usepackage[utf8]{inputenc}
\usepackage{graphicx}
\usepackage{mathtools}

\usepackage{color}

\newcommand*{\mat}[1]{\boldsymbol{#1}}
\newcommand*{\coord}[1]{\mathbf{#1}}
\newcommand*{\veck}{\coord{k}}
\newcommand*{\vecr}{\coord{r}}

\newcommand*{\vecx}{\coord{x}}

\newcommand*{\binteg}[3]{\int^{\mathrlap{#3}}_{\mathrlap{#2}}\ud{#1}\,}
\newcommand*{\integ}[1]{\int\!\!\!\:\ud{#1}\:}
\newcommand*{\iinteg}[2]{\integ{#1}\!\!\!\integ{#2}}

\newcommand*{\crea}[1]{\hat{#1}^{\dagger}}
\newcommand*{\anni}[1]{\hat{#1}^{\vphantom{\dagger}}}

\newcommand*{\leftnabla}{%
  \raisebox{1.5ex}{\makebox[-0.5pt][l]{$\scriptstyle \leftharpoonup$}}\nabla}
\newcommand*{\rightnabla}{%
  \raisebox{1.5ex}{\makebox[-0.3pt][l]{$\scriptstyle \rightharpoonup$}}\nabla}
\newcommand*{\leftdu}{%
  \raisebox{1.5ex}{\makebox[-0.5pt][l]{$\scriptstyle \leftharpoonup$}}\du}
\newcommand*{\rightdu}{%
  \raisebox{1.5ex}{\makebox[-0.3pt][l]{$\scriptstyle \rightharpoonup$}}\du}

\newcommand{\brakket}[3]{\langle{#1}|{#2}|{#3}\rangle}

\DeclareMathOperator{\Fourier}{\mathcal{F}}

\DeclareMathOperator{\BesselK}{\mathrm{K}}
\DeclareMathOperator{\Order}{\mathcal{O}}
\DeclareMathOperator{\sgn}{sgn}

\newcommand*{\abs}[1]{\lvert#1\rvert}

\newcommand*{\du}{\partial}
\newcommand*{\e}{\textrm{e}}
\newcommand*{\eqspace}{\hphantom{{}={}}}

\newcommand*{\Ecal}{\mathcal{E}}
\newcommand*{\half}{\frac{1}{2}}

\newcommand*{\isDefinedAs}{\coloneqq}
\newcommand{\norm}[1]{\lVert#1\rVert}
\newcommand{\Bgnorm}[1]{\Bigl\lVert#1\Bigr\rVert}
\newcommand*{\Reals}{\mathbb{R}}
\newcommand*{\ud}{\mathrm{d}}

\newcommand*{\figdir}{.}

\allowdisplaybreaks[4]

\begin{document}

\title{Compact two-electron wave function for bond dissociation and Van der Waals interactions: A~natural amplitude assessment}
\author{Klaas J. H. Giesbertz}
\affiliation{Theoretical Chemistry, Faculty of Exact Sciences, VU University, De Boelelaan 1083, 1081 HV Amsterdam, The Netherlands}
\author{Robert van Leeuwen}
\affiliation{Department of Physics, Nanoscience Center, University of Jyväskylä, P.O. Box 35, 40014 Jyväskylä, Survontie 9, Jyväskylä, Finland}

\date{\today}

\begin{abstract}
Electron correlations in molecules can be divided in short range dynamical correlations, long range Van der Waals type interactions and near degeneracy static correlations. In this work we analyze for a one-dimensional model of a two-electron system how these three types of correlations can be incorporated in a simple wave function of restricted functional form consisting of an orbital product multiplied by a single correlation function $f(r_{12})$ depending on the interelectronic distance $r_{12}$. Since the three types of correlations mentioned lead to different signatures in terms of the natural orbital (NO) amplitudes in two-electron systems we make an analysis of the wave function in terms of the NO amplitudes for a model system of a diatomic molecule. In our numerical implementation we fully optimize the orbitals and the correlation function on a spatial grid without restrictions on their functional form. Due to this particular form of the wave function, we can prove that none of the amplitudes vanishes and moreover that it displays a distinct sign pattern and a series of avoided crossings as a function of the bond distance in agreement with the exact solution. This shows that the wave function Ansatz correctly incorporates the long range Van der Waals interactions. We further show that the approximate wave function gives an excellent binding curve and is able to describe static correlations. We show that in order to do this the correlation function $f(r_{12})$ needs to diverge for large $r_{12}$ at large internuclear distances while for shorter bond distances it increases as a function of $r_{12}$ to a maximum value after which it decays exponentially. We further give a physical interpretation of this behavior.
\end{abstract}

\maketitle

\section{Introduction}
\label{sec:introduction}

The efficient description of electronic correlations is a key problem in electronic structure theory. 
The electronic correlations can be divided in several types. To describe the dissociation of molecules
static or near-degeneracy correlations play an important role. This type of correlations is typically taken
into account by a few well chosen terms in a configuration interaction (CI) expansion of the wave function
in terms of Slater determinants~\cite{Slater1929} in terms of Hartree-Fock (HF) orbitals~\cite{Hartree1928, Fock1930}. Also long range correlations, such as
Van der Waals interactions, can be described with a few well-chosen configurations~\cite{GritsenkoBaerends2006, MentelShengGritsenko2012, ShengMentelGritsenko2013}.
The remaining correlations are the short range correlations which describe the interelectronic cusp
\begin{align}\label{eq:cuspCondition}
\Psi(r_{12} \to 0) = \Psi(r_{12} = 0)\left(1 + \half r_{12} + \dotsb\right),
\end{align}
where $r_{12} \isDefinedAs \abs{\vecr_1 - \vecr_2}$. The cusp is due to the Coulomb interaction between the electrons which becomes infinite when the electrons approach each other, $r_{12} \to 0$. To compensate for this infinite interaction energy the wave function needs to have a kink at this point, which gives an infinite kinetic energy which exactly compensates the divergence of the Coulomb interaction~\cite{Lowdin1954, Kato1957, PackBrown1966, FournaisHoffmann-OstenhofHoffmann-Ostenhof2005}.
The description of this cusp requires the inclusion of a large number of Slater determinants in a CI expansion.

In two recent papers~\cite{GiesbertzLeeuwen2013a, GiesbertzLeeuwen2013b} we found that the three types of correlation mentioned lead to different signatures in the natural orbital (NO) occupation numbers and amplitudes. 
The NOs, $\varphi_k(\vecx)$, and their occupations, $n_k$, are defined as the eigenfunctions and eigenvalues respectively of the one-body reduced density matrix (1RDM)
\begin{align*}
\gamma(\vecx,\vecx') \isDefinedAs \brakket{\Psi}{\crea{\psi}(\vecx')\anni{\psi}(\vecx)}{\Psi}
= \sum_kn_k\varphi_k(\vecx)\varphi_k^*(\vecx'),
\end{align*}
where $\vecx \isDefinedAs \vecr\sigma$ is a combined space and spin coordinate. 
The presence of the Coulomb cusp leads to a power law decay rate of the NO occupation
numbers $n_k \sim k^{-p}$ where $p>0$ when they are ordered as a descending series~\cite{GiesbertzLeeuwen2013a} while the description of strong correlations is signaled 
by strong deviations of the  NO occupations
from one and zero. The long range Van der Waals type interactions are responsible for the sign pattern of the NO amplitudes in two-electron systems~\cite{GiesbertzLeeuwen2013b}. 
Two-electron systems have the special property that the wave function can be diagonalized to yield one-particle orbitals. In particular for the singlet case, the spatial part of the wave function is symmetric and can be brought to diagonal form as~\cite{LowdinShull1956}
\begin{align*}
\Psi(\vecr_1,\vecr_2) = \sum_k c_k \,\varphi_k(\vecr_1)\varphi_k (\vecr_2).
\end{align*}
It is not difficult to show that the eigenfunctions are NOs and that the expansion coefficients, also called the natural amplitudes, are related to the occupation numbers as $n_k = \abs{c_k}^2$. Since the NOs are eigenfunctions of the singlet two-electron wave function, they satisfy the following integral equation
\begin{align}
\integ{\vecr_2}\Psi(\vecr_1,\vecr_2)\varphi_k(\vecr_2) = c_k\,\varphi_k(\vecr_1).
\label{eq:ck}
\end{align}
The sign pattern of the largest amplitudes $c_k$ is determined by the long-range properties of the two-body interaction which in turn determines the asymptotic decay of the
two-electron wave function. In diatomic molecules the amplitudes display a distinct behavior as a function of the bond distance $R$. The curves $c_k (R)$ never cross zero but
display a series of avoided crossings at which amplitudes of one sign get small and amplitudes of the opposite sign grow. The Van der Waals limit is characterized by the
fact that several amplitudes acquire significant positive values for large values of $R$~\cite{ShengMentelGritsenko2013, GiesbertzLeeuwen2013b}.  This is a direct consequence of the induced dipole-dipole and higher order interactions~\cite{ShengMentelGritsenko2013,CioslowskiPernal2006}.

Now that we established the relation between various correlation effects and the properties of the NO occupations and amplitudes we ask for the simplest
approximate compact wave function that displays all these features in order to facilitate a transparent physical interpretation.
 We therefore demand from our approximate wave function that it has the following
properties:
\begin{enumerate}
\item It has a compact representation (restricted functional form)
\item It correctly describes the short range correlations between the electrons
\item It displays the correct sign pattern and avoided crossings of the NO amplitudes
\item It produces the correct bonding curves and dissociation limit
\end{enumerate}
The first and second point can be taken into account by using directly the electronic distance $r_{ij}=|\vecr_i-\vecr_j|$ between the
electrons as a variable. We will consider the Ansatz
\begin{align}
\label{eq:PsiF12Phi}
\Psi(\vecx_1,\vecx_2) = f(\vecr_{12})\Phi(\vecx_1,\vecx_2),
\end{align}%
where $\vecx \isDefinedAs \vecr\sigma$ is a combined space-spin coordinate, $f(\vecr_{12})$ is a function which ensures the correct asymptotic behavior of $\Psi$ for $r_{12} \to 0$ and $\Phi$ is an orbital based Ansatz for the wave function, e.g.\ a Slater determinant, a restricted HF expression or some limited CI expansion. The function $f(r_{12})$ not only has a clear physical interpretation but also
obviates the need of a large CI expansion. This is exactly the reason why wave functions of the form in Eq.~\eqref{eq:PsiF12Phi} have received considerable attention in the study of many-electron
systems~\cite{KlopperManbyTen-No2006, KongBischoffValeev2011}. 
In the quantum chemistry community, initially only the linear term from~\eqref{eq:cuspCondition} was taken into account, $f(r_{12}) = 1 + \half r_{12}$~\cite{Hylleraas1929, Kutzelnigg1985}. The main reason for this simple form is to make the evaluation of the integrals not too overly complicated. However, the main disadvantage is that the simple linear form does not have a proper asymptotic behavior, so large basis sets are still required to make up for this deficiency~\cite{TewKlopper2005}. To remedy this deficiency, simple alternative correlation functions have been put forward, e.g.\ $f(r_{12}) = \e^{-\zeta r_{12}}$~\cite{GreenStephensKolchin1959, TewKlopper2005, KlopperManbyTen-No2006} and lead to a greatly enhanced convergence with respect to the basis set. Within the variational quantum Monte Carlo and diffusion quantum Monte Carlo more involved correlation functions have already been in use for a long time, since the required integrals are evaluated by the Metropolis integration technique, so one does not have to be concerned about the complexity of the correlation function. Usually, the correlation function in variational and diffusion quantum Monte Carlo is written as an exponential, $f(r_{12}) = \e^{u(r_{12})}$, and is called the Jastrow factor~\cite{Dingle1949, Jastrow1955, FoulkesMitasNeeds2001}. Unfortunately, the simultaneous optimization of both the Jastrow factor and the orbital expansion $\Phi$ is not straightforward~\cite{SchautzFilippi2004}.
Not too much is known about the behavior of the correlation function $f(r_{12})$, except that it is a possibly diverging function~\cite{BaberHasse1937, GreenStephensKolchin1959} for $r_{12} \rightarrow \infty$ and that it should behave linearly the small $r_{12}$ limit such that the total wave function satisfies the cusp condition~\eqref{eq:cuspCondition}.

A wave function of the form~\eqref{eq:PsiF12Phi} has been considered before~\cite{VarganovMartineza2010}, though the correlation function has never been optimized in a self-consistent manner.
In this article we want to optimize both $\Phi$ and the correlation function simultaneously to gain a better understanding of the required features of the correlation function. 
In particular we want to study the behavior of this function as function of the bond distance in a diatomic molecule in order to see if our Ansatz wave function is able to describe
molecular dissociation correctly.
To simplify the computations, we have chosen to consider only two electrons coupled to a singlet, so we only need to consider the spatial part of the wave function which should be symmetric. As an additional simplification, we use the restricted HF approximation for the orbital part $\Phi$, so we will consider the following approximate wave function
\begin{align}\label{eq:explCorWave3D}
\Psi^{\phi^2\!f} (\vecr_1,\vecr_2) = \phi(\vecr_1)\phi(\vecr_2)f(\vecr_{12}),
\end{align}
where $\vecr_{12} \isDefinedAs \vecr_1 - \vecr_2$. Although this wave function has only a single reference at its core and the correlation function was originally included to handle the electron-electron cusp~\eqref{eq:cuspCondition}, there is no reason to assume that the correlation function could not handle other types of correlation. 
It is important to realize that there are only two electrons, so the correlation can be fully adjusted to describe the correlations between this electron pair.
In particular, we will demonstrate that a fully optimized $f(\vecr_{12})$ and $\phi(\vecr)$ can actually capture all the required correlation for a dissociating H$_2$ molecule. In particular we can argue that requirement 3 in our list is likely to be satisfied since 
for our Ansatz we can prove that none of the NO amplitudes will vanish such that sign changes in the amplitudes can only occur by avoided crossings. This can be seen as follows~\cite{GiesbertzLeeuwen2013a}. From Eq.(\ref{eq:ck})
we see that $c_k=0$ is only possible for our Ansatz when
\begin{align}\label{eq:zeroNOcond}
\integ{\vecr_2}f(\vecr_1-\vecr_2)\chi_i(\vecr_2) = 0,
\end{align}
where $\chi_i(\vecr) \isDefinedAs \phi(\vecr)\varphi_i(\vecr)$. Taking the Fourier transform $\Fourier$ we can turn the convolution into a normal product
\begin{align*}
\Fourier[f](\veck) \cdot \Fourier[\chi_i](\veck) = 0.
\end{align*}
If $\Fourier[f](\veck) \neq 0$ almost everywhere, this implies that $\Fourier[\chi_i](\veck) = 0$. Since $\phi(\vecr) > 0$, this can only be the case when $\varphi_i(\vecr) = 0$. However, this is not a normalizable function, so no $c_i = 0$ exist if $\Fourier[f](\veck) \neq 0$ almost everywhere. Conversely, if $\Fourier[f](\veck) = 0$ on some finite interval, we can readily construct an unoccupied NO by choosing $\Fourier[\chi_i](\veck) \neq 0$ on this interval and zero outside. Fourier transforming back to real space and dividing out the orbital $\phi(\vecr)$, we have constructed an NO with zero occupation number. We will use a modified version of this theorem later that our numerically obtained Ansatz wave function does not have any unoccupied NOs.

The article is structured as follows. In Sec.~\ref{sec:theory} we will derive differential equations for the orbital, $\phi(\vecr)$, and the correlation function, $f(\vecr_{12})$, from the variational principle. To make the presentation and numerics as simple as possible we restrict ourselves to
a one-dimensional model description of the diatomic molecule. We give a basic explanation how to construct a numerical solution. The details are quite involved and can be found in the Appendix~\ref{ap:technics}.
In Sec.~\ref{sec:results} we will present our results from the optimization of $\Psi$ 
and discuss the properties of the correlation function $f(r_{12})$ as a function of the bond distance.
We then direct our attention to the issue of unoccupied NOs in Sec.~\ref{sec:occVanish} and the properties of the NO amplitudes. We show that the
NO amplitudes display a series of avoided crossings as a function of the bond distance in close resemblance to the exact wave function and 
conclude that our Ansatz can properly account for the correct long-range structure in the wave function which is necessary for a good description of Van der Waals interactions. Our final conclusions are presented in Sec.~\ref{sec:conclusion}.

\section{Theory}
\label{sec:theory}
We will first derive differential equations that the orbital, $\phi(\vecr)$, and correlation function, $f(\vecr_{12})$ have to satisfy for the energy to be minimal. Consider a general Hamiltonian of the form
\begin{align*}
\hat{H} = -\half\bigl(\nabla_1^2 + \nabla_2^2\bigr) + V(\vecr_1,\vecr_2),
\end{align*}
where all the potentials have been combined into
\begin{align*}
V(\vecr_1,\vecr_2) = v(\vecr_1) + v(\vecr_2) + w(r_{12}).
\end{align*}
The energy for the explicitly correlated ansatz~\eqref{eq:explCorWave3D} can be expressed as
\begin{align*}
E = \frac{\Ecal[\phi,f]}{N[\phi,f]},
\end{align*}
where we introduced
\begin{align*}
\Ecal &\coloneqq \iinteg{\vecr_1}{\vecr_2}\phi(\vecr_1)\phi(\vecr_2)f(\vecr_{12})
\hat{H}\phi(\vecr_1)\phi(\vecr_2)f(\vecr_{12}), \\
N &\coloneqq \iinteg{\vecr_1}{\vecr_2}\phi^2(\vecr_1)\phi^2(\vecr_2)f^2(\vecr_{12}).
\end{align*}
To find the ground state, we require the energy functional to be stationary with respect to variations in the orbital and the correlation function
\begin{align}\label{eq:Egradient}
\begin{split}
0 = \frac{\delta E}{\delta \phi(\vecr)} 
= \frac{1}{N}\biggl(\frac{\delta \Ecal}{\delta \phi(\vecr)} - E\frac{\delta N}{\delta \phi(\vecr)}\biggr), \\
0 = \frac{\delta E}{\delta f(\vecr)} 
= \frac{1}{N}\biggl(\frac{\delta \Ecal}{\delta f(\vecr)} - E\frac{\delta N}{\delta f(\vecr)}\biggr).
\end{split}
\end{align}
Working out the functional derivatives, the stationarity equations for the orbital and correlation function can be cast in the form of effective Schrödinger equations
\begin{subequations}\label{eq:stationary}
\begin{align}
\biggl(-\half\nabla^2 - \mat{\alpha}(\vecr)\cdot\nabla + \beta(\vecr)\biggr)\phi(\vecr) 
&= E\,\phi(\vecr), \\
\label{eq:fStat}
\bigl(-\nabla^2 - \mat{\mu}(\vecr)\cdot\nabla + \nu(\vecr)\bigr)f(\vecr) &= E\,f(\vecr),
\end{align}
\end{subequations}
where we introduced the following quantities
\begin{align}\label{eq:statIntegrals}
\mat{\alpha}(\vecr) &\coloneqq \frac{\integ{\vecr'}\phi^2(\vecr')
f(\vecr-\vecr')\nabla_{\vecr} f(\vecr-\vecr')}
{\integ{\vecr'}\phi^2(\vecr')f^2(\vecr-\vecr')}, \notag \\
\beta(\vecr) &\coloneqq \frac{\integ{\vecr'}\phi(\vecr')f(\vecr-\vecr')
\hat{H}(\vecr,\vecr') \phi(\vecr')f(\vecr-\vecr')}{\integ{\vecr'}\phi^2(\vecr')f^2(\vecr-\vecr')}, \notag \\
\mat{\mu}(\vecr) &\coloneqq \frac{\integ{\vecr'}\phi^2(\vecr')
\bigl[\phi(\vecr_1)\nabla_1 \phi(\vecr_1)\bigl]_{\vecr_1=\vecr'+\vecr}}
{\integ{\vecr'}\phi^2(\vecr')\phi^2(\vecr'+\vecr)} - {} \\*
&\eqspace
\frac{\integ{\vecr'}\phi^2(\vecr')
\bigl[\phi(\vecr_1)\nabla_1 \phi(\vecr_1)\bigl]_{\vecr_1=\vecr'-\vecr}}
{\integ{\vecr'}\phi^2(\vecr')\phi^2(\vecr'-\vecr)}, \notag \\
\nu(\vecr) &\coloneqq \frac{\integ{\vecr'}\bigl[\phi(\vecr')\phi(\vecr_1)
\hat{H}(\vecr_1,\vecr')\phi(\vecr_1)\phi(\vecr')\bigr]_{\vecr_1=\vecr+\vecr'}}
{\integ{\vecr'}\phi^2(\vecr')\phi^2(\vecr+\vecr')}. \notag
\end{align}
First note that the function $\mat{\mu}(\vecr)$ and $\nu(\vecr)$ only depend on the orbital and not on the correlation function. Thus the differential equation for the correlation function~\eqref{eq:fStat} is effectively an eigenvalue equation for $f(\vecr)$. So given the orbital, we can calculate the corresponding correlation function by solving this eigenvalue equation. It is particularly interesting to consider the HF orbital and then to solve for the correlation function and see how much the description of the ground state of H$_2$ improves without any orbital relaxation. As we will show later in Sec.~\ref{sec:results}, this will already bring in the major part of the required static correlation to correctly dissociate the hydrogen molecule.

Our goal is to find a complete stationary state (hopefully the ground state), so these equations have to be solved together to self-consistency. However, this set of equations is rather intimidating and can hardly be regarded as a simplification of the original problem (of solving the Schrödinger equation). In the case of the helium atom, the spherical symmetry reduces the complexity of the problem, since the functions have only a radial part, $\phi(r)$ and $f(r_{12})$, so only one dimensional functions have to be found. The case of the helium atom has already been considered before by Green et al.~\cite{GreenStephensKolchin1959}. Although the functions are now only one dimensional, the differential equations resulting from their variational optimization were still too complicated to be solved completely, mainly due to a lack of computational resources at that time. As a simplification, Green et al.\ limited the orbital to the form $\phi(r) = \e^{-Zr}$ and optimized the exponent $Z$. The full correlation function could now be solved from its differential equation. Most interestingly, they found that $f(r_{12})$ diverges exponentially for large values of $r_{12}$.

Since we are interested in the question whether the correlation function has the required flexibility to handle the strong correlation effects present in a dissociating chemical bond, we should consider the hydrogen molecule. Because the most interesting physics occurs along the bond axis, we will limit ourselves to a 1D model of the hydrogen atom which corresponds to the electronic Hamiltonian
\begin{align*}
\hat{H} = -\half\left(\frac{\du^2}{\du x_1^2} + \frac{\du^2}{\du x_2^2}\right) + v(x_1) + v(x_2) + w(x_{12}),
\end{align*}
where $v(x)$ is the potential due to the nuclei and $w(x_{12})$ denotes the interaction between the electrons.
In 1D the Coulomb potential becomes too singular and the Hamiltonian becomes unbounded from below. 
Therefore, the Coulomb singularity needs to be removed in 1D. Because the potential is not allowed to diverge anymore in 1D, the 1D wave function will not have a cusp at the coalescence points. In our work we have chosen to use soft Coulomb potentials to remove the singularity at $r = 0$, while still retaining the proper $1/r$ behavior for $r \to \infty$~\cite{JavanainenEberlySu1988, LappasLeeuwen1998, HelbigFuksCasula2011, FuksRubioMaitra2011, dahlen2002, kreibich2001, kreibich2002, kreibich2004}. Though the 1D wave function will not have a cusp like its 3D counter part, the qualitative physical behavior will be the same, since they are mainly dictated by the Coulomb tail.
So for the hydrogen molecule we use
\begin{align}
\label{eq:nucPot}
v(x) &= \frac{\lambda_v}{\sqrt{(x-\rho)^2 + \alpha_v^2}} + 
\frac{\lambda_v}{\sqrt{(x+\rho)^2 + \alpha_v^2}}, \\
w(x) &= \frac{\lambda_w}{\sqrt{x^2 + \alpha_w^2}}, \notag
\end{align}
where $2\rho$ is the distance between the nuclei and $\alpha_v$ and $\alpha_w$ are the softening parameters, cf.\ with $\alpha_v = \alpha_w = 0$ we recover the full Coulomb potential. In our calculations we have simply set $\alpha_v = \alpha_w = 1$. These parameters give an equilibrium bond-length of $R_e \approx 1.5$ Bohr which is quite close to the equilibirum bond length of H$_2$ in 3D. For the charges we have used the normal electron and proton charges, so $\lambda_v = -1$ and $\lambda_w = 1$. Note that in principle we could retain the full Coulomb potential for the interaction, since it is repulsive. However, it is more physical to soften this potential as well, otherwise the electrons would not be able to pass each other and the wave function would vanish at $x_{12} = 0$.

The explicitly correlated Ansatz for the 1D hydrogen molecule becomes
\begin{align}\label{eq:explCorWave1D}
\Psi^{\phi^2\!f}(x_1,x_2) = \phi(x_1)\phi(x_2)f(x_{12}),
\end{align}
where $x_{12} \isDefinedAs x_1 - x_2$.
The correlation function in 1D will not have the cusp that its 3D counterpart has, since the Coulomb singularity needed to be removed. Apart from the lack of a cusp, it is expected that the 1D correlation function will be quite similar to its 3D cousin. Since we only modified the short-range part of the interaction, only the short-range part of the correlation function should be affected.

The reduced complexity now ensures that a fully self-consistent solution is feasible, though still a formidable task.
We use the following approach for the optimization. First we set $f = 1$ and only optimize the orbital, which corresponds to a Hartree--Fock (HF) optimization. Next we only solve the effective Schrödinger equation for the correlation function~\eqref{eq:fStat} using the HF orbitals obtained in the the previous step. Since, the (vector)potentials $\mu(x)$ and $\nu(x)$~\eqref{eq:statIntegrals} do not depend explicitly on the correlation function, this equation~\eqref{eq:fStat} is an effective eigenvalue equation for the correlation function and can readily be solved with standard linear algebra. This partial optimized solution already provides a significant improvement over the bare HF wave function, so will also be under consideration in this work. In the next step we use non-linear algorithms to optimize the energy with respect to the orbital and the correlation function simultaneously. The HF orbital and its corresponding correlation function obtained in the previous step can be used as an initial guess for short bond distances. For elongated bond distances this initial guess quickly deteriorates and one can better use the optimized orbital and correlation function from a similar bond distance. The details about the numerical challenges are quite involved and have been deferred to the appendix~\ref{ap:technics}. We rather like to focus our findings in the next section (Sec.~\ref{sec:results}).

\section{Correlation function and bond dissociation}
\label{sec:results}

\subsection{Optimizing the correlation function for fixed orbital}

\begin{figure}[t]
  \includegraphics[width=\columnwidth]{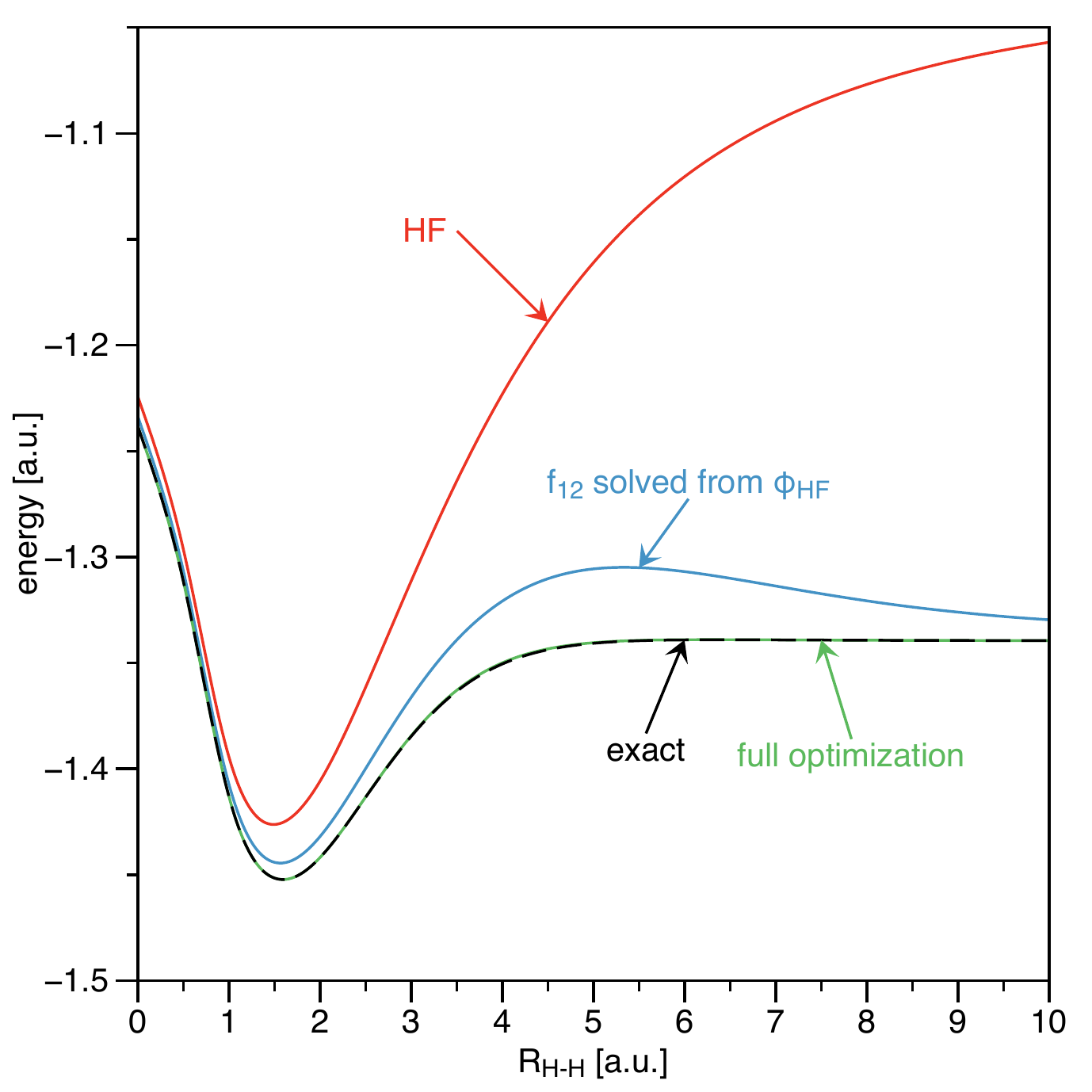}
  \caption{The total energy as a function of the bond distance, $R_{\text{H--H}}$, at different stages of the optimization. The exact result is shown by the black dashes as a reference.}
  \label{fig:energies}
\end{figure}

\begin{figure}[t]
  \includegraphics[width=\columnwidth]{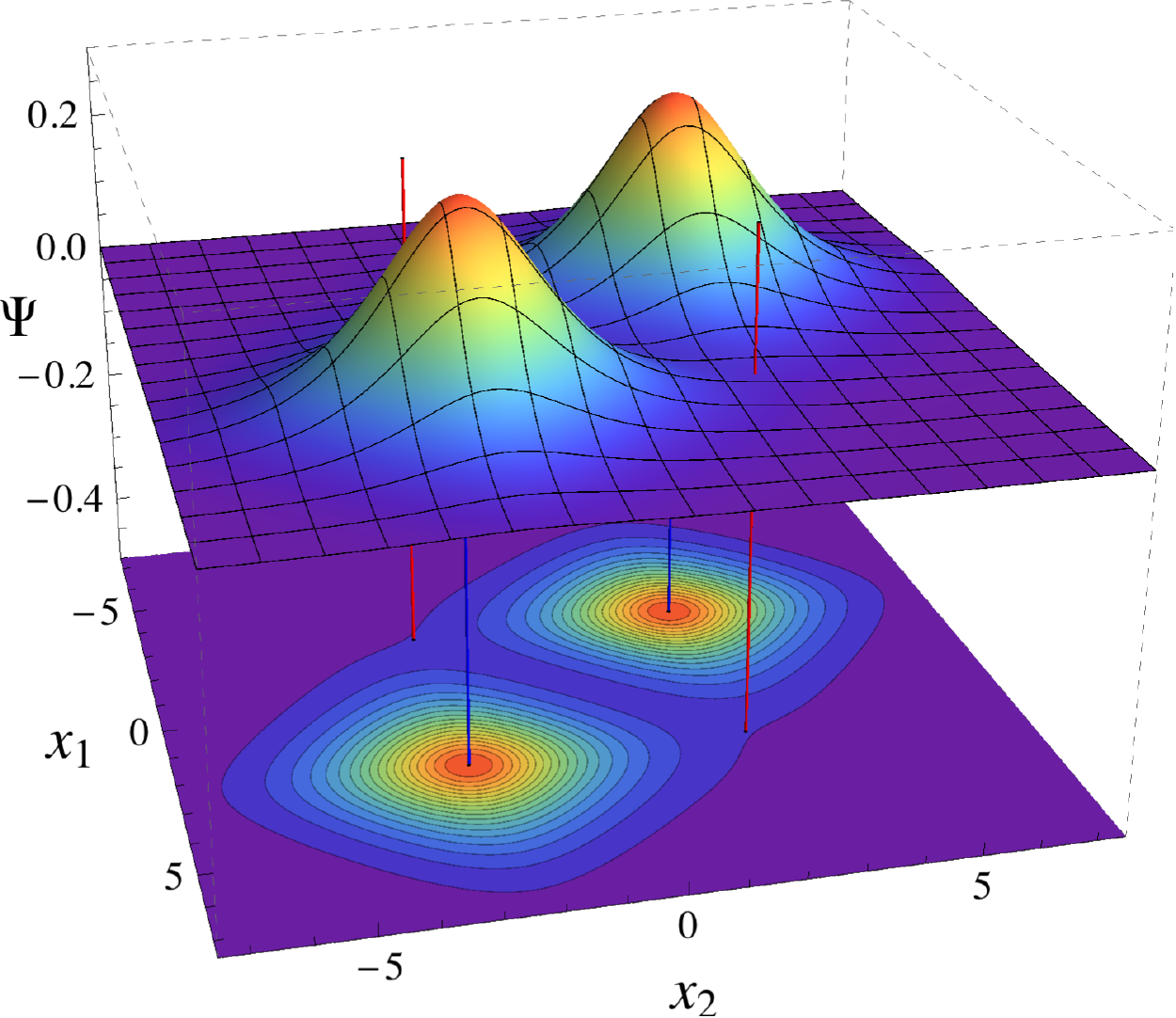}
  \caption{The numerically exact wave function in atomic units for an interatomic distance of $R_{\text{H--H}} =$ 5.0 Bohr. The positions of the nuclei are indicated by the vertical blue and red lines. The blue lines correspond to covalent configurations and the red lines indicate the ionic configurations.}
  \label{fig:exactWaveAt5Bohr}
\end{figure}

In this section we will discuss the results from the different stages of the optimization of the explicitly correlated wave function~\eqref{eq:explCorWave1D}. In Fig.~\ref{fig:energies} we show the results for the total energy for HF, the HF orbital combined with the corresponding correlation function from its differential equation~\eqref{eq:fStat} and the fully optimized explicitly correlated wave function (in decreasing order of energy). These results are compared with a numerically exact calculation, where we discretized the wave function in the $X_{12} \coloneqq x_1+x_2$ and $x_{12} \coloneqq x_1-x_2$ direction. The grid in the center-of-mass coordinate system makes it easier to utilize the full symmetry of the wave function. Since $\Psi(X_{12},x_{12}) = \Psi(X_{12},-x_{12})$ (identical particles) and $\Psi(X_{12},x_{12}) = \Psi(-X_{12},-x_{12})$ (mirror symmetry of H$_2$), one only has to take $X_{12} \in [0,\infty)$ and $x_{12} \in [0,\infty)$ into account. We used the same spline-machinery as exposed in the Appendix~\ref{ap:technics} with $b=15 + R_{\text{H--H}}$, $c=b+5$ and $20c$ grid points in both directions ($X_{12}$ and $x_{12}$), which gave sufficiently converged results. The numerically exact wave function for $R_{\text{H--H}} = 5.0$ Bohr is shown in Fig.~\ref{fig:exactWaveAt5Bohr}.

\begin{figure}[t]
  \includegraphics[width=0.5\columnwidth]{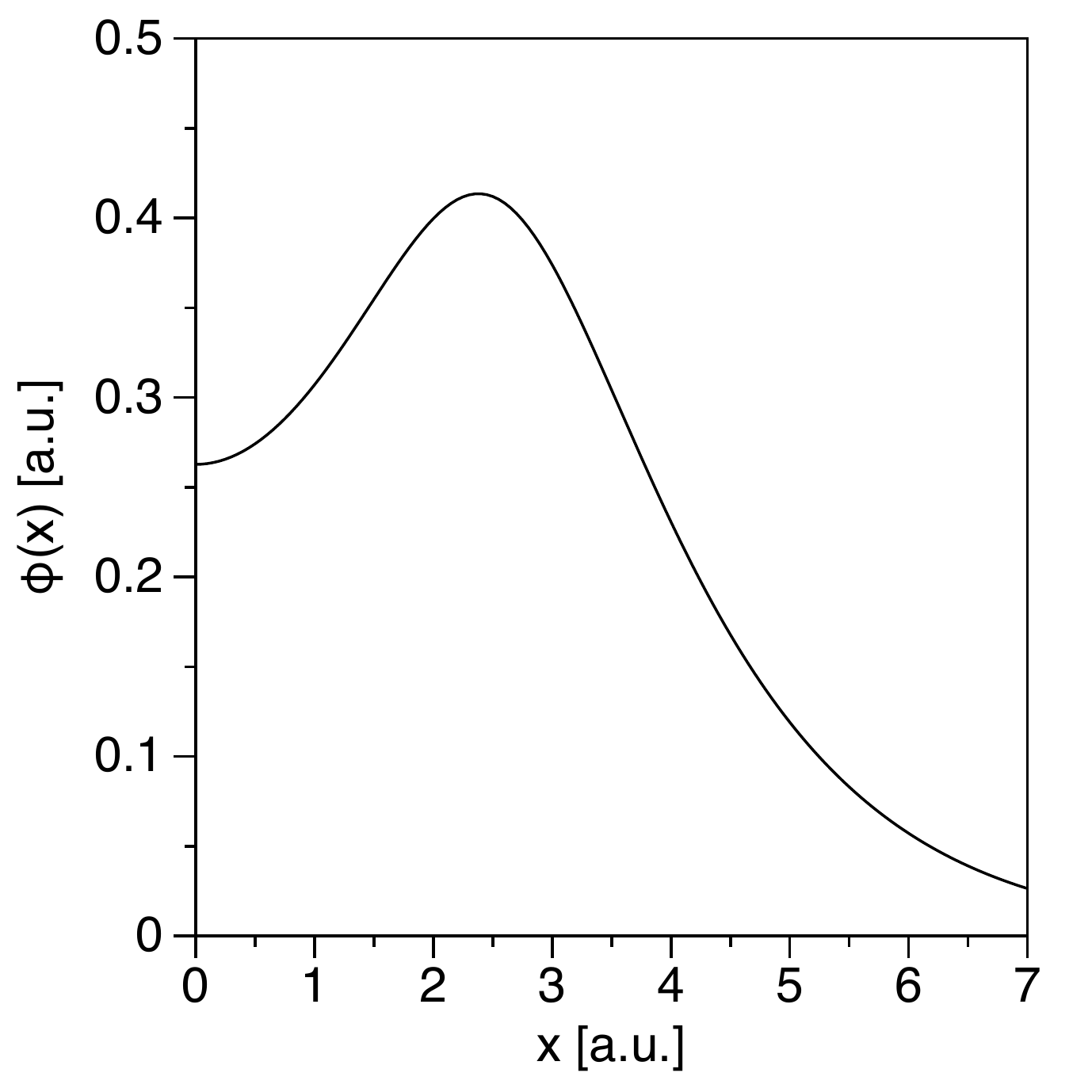}%
  \includegraphics[width=0.5\columnwidth]{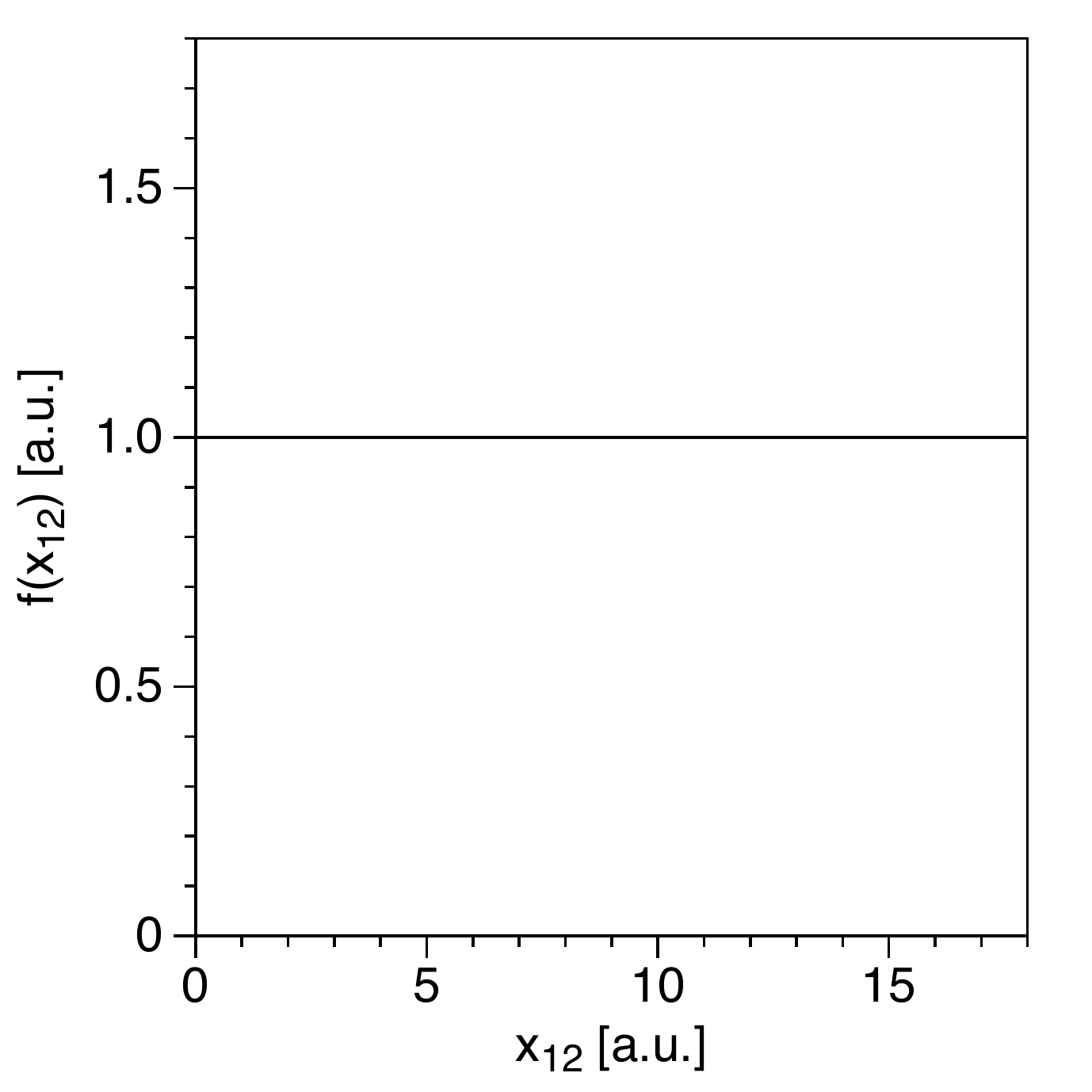}
  \includegraphics[width=\columnwidth]{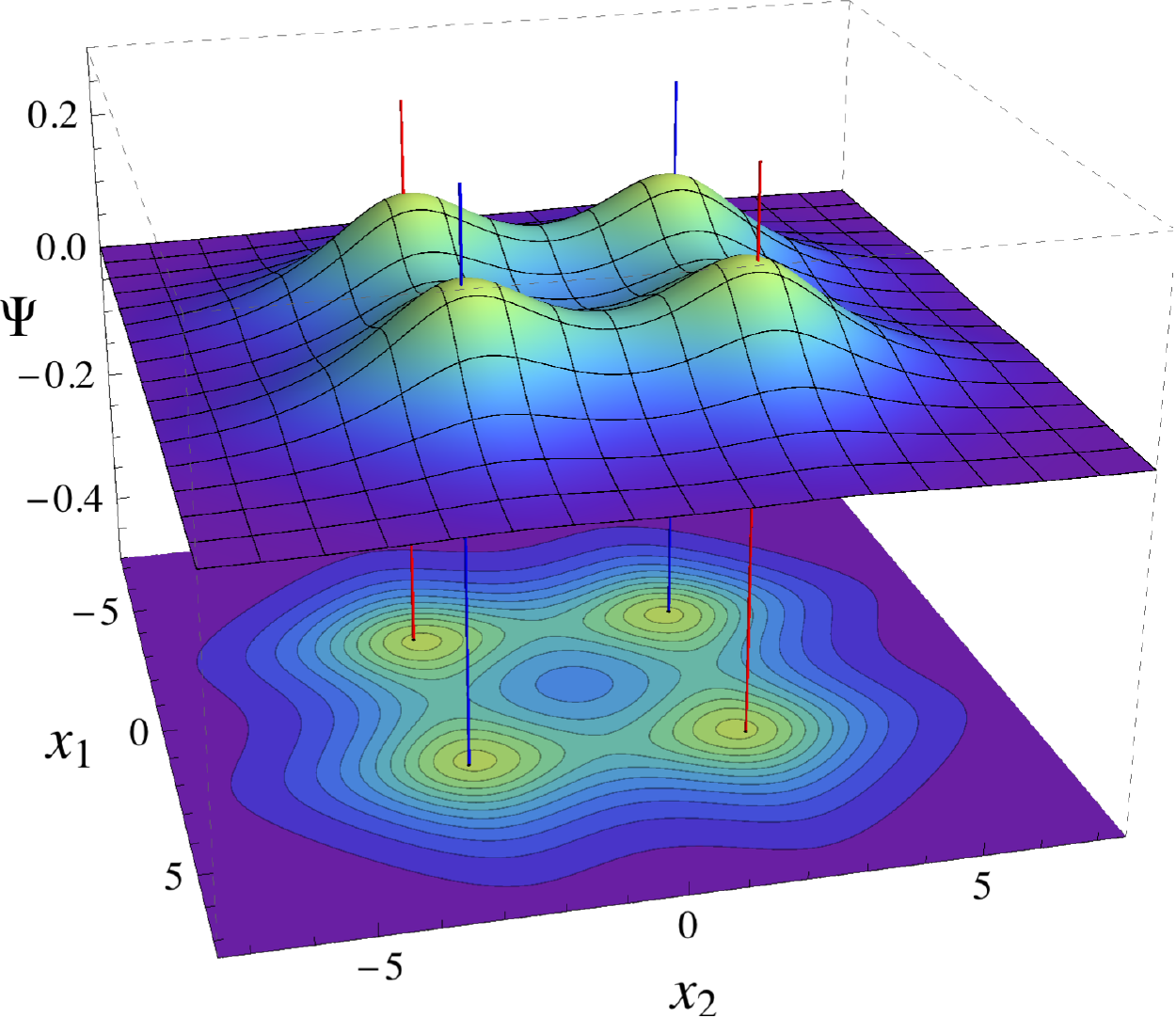}
  \caption{In the upper-left panel the HF orbital, $\phi(x)$, as a function of the distance from the bond mid-point in the upper right panel the HF correlation function, $f(x) = 1$, and in the lower panel the full HF wave function $\Psi^{\phi^2\!f}(x_1,x_2) = \phi(x_1)\phi(x_2)f(x_{12})$ in atomic units at the internuclear distance of $R_{\text{H--H}} = 5.0$ Bohr. The positions of the nuclei are indicated by the vertical blue and red lines again as in Fig.~\ref{fig:exactWaveAt5Bohr}.}
  \label{fig:HFat5Bohr}
\end{figure}

\begin{figure}[t]
  \includegraphics[width=0.5\columnwidth]{"\figdir/HForbAt5Bohr"}%
  \includegraphics[width=0.5\columnwidth]{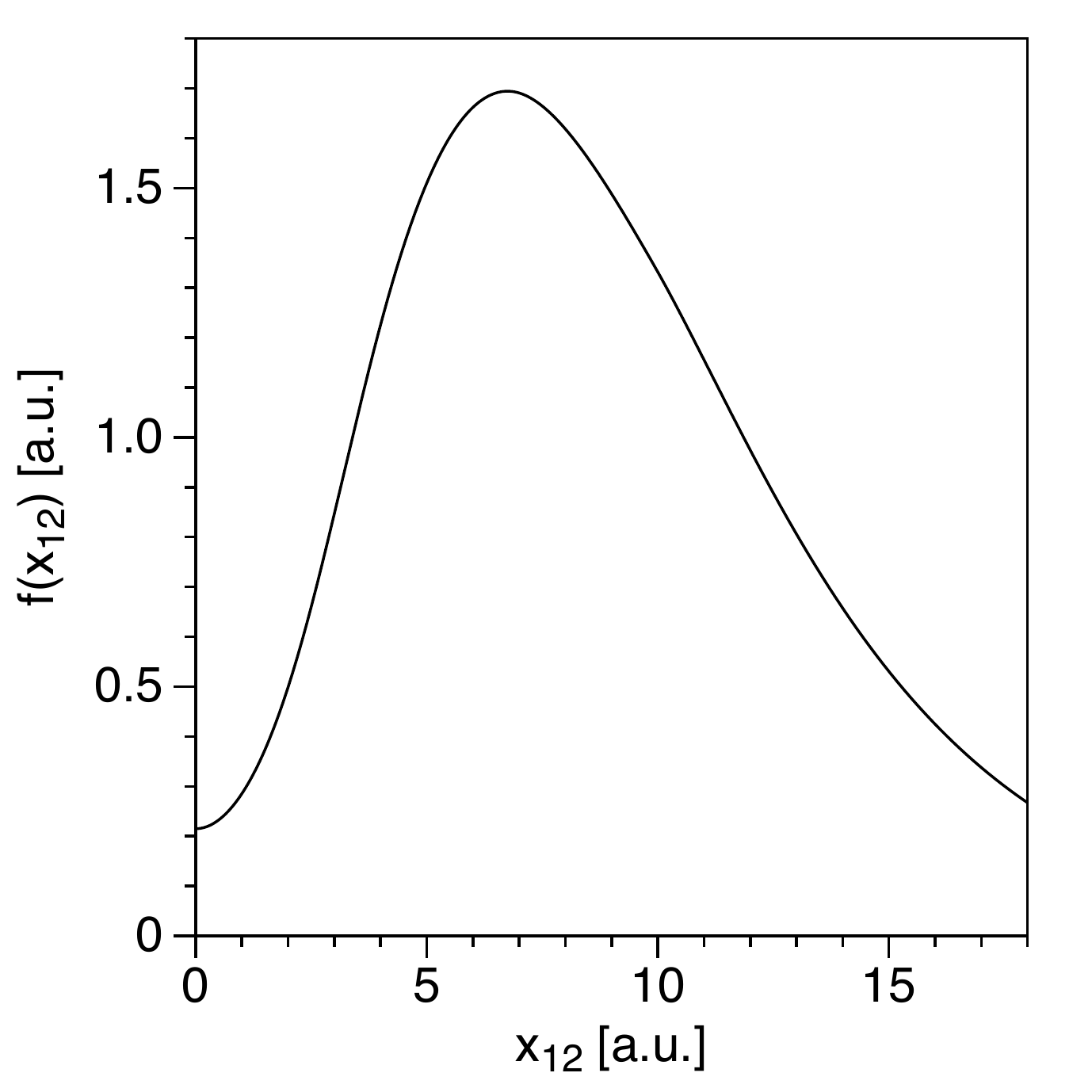}
  \includegraphics[width=\columnwidth]{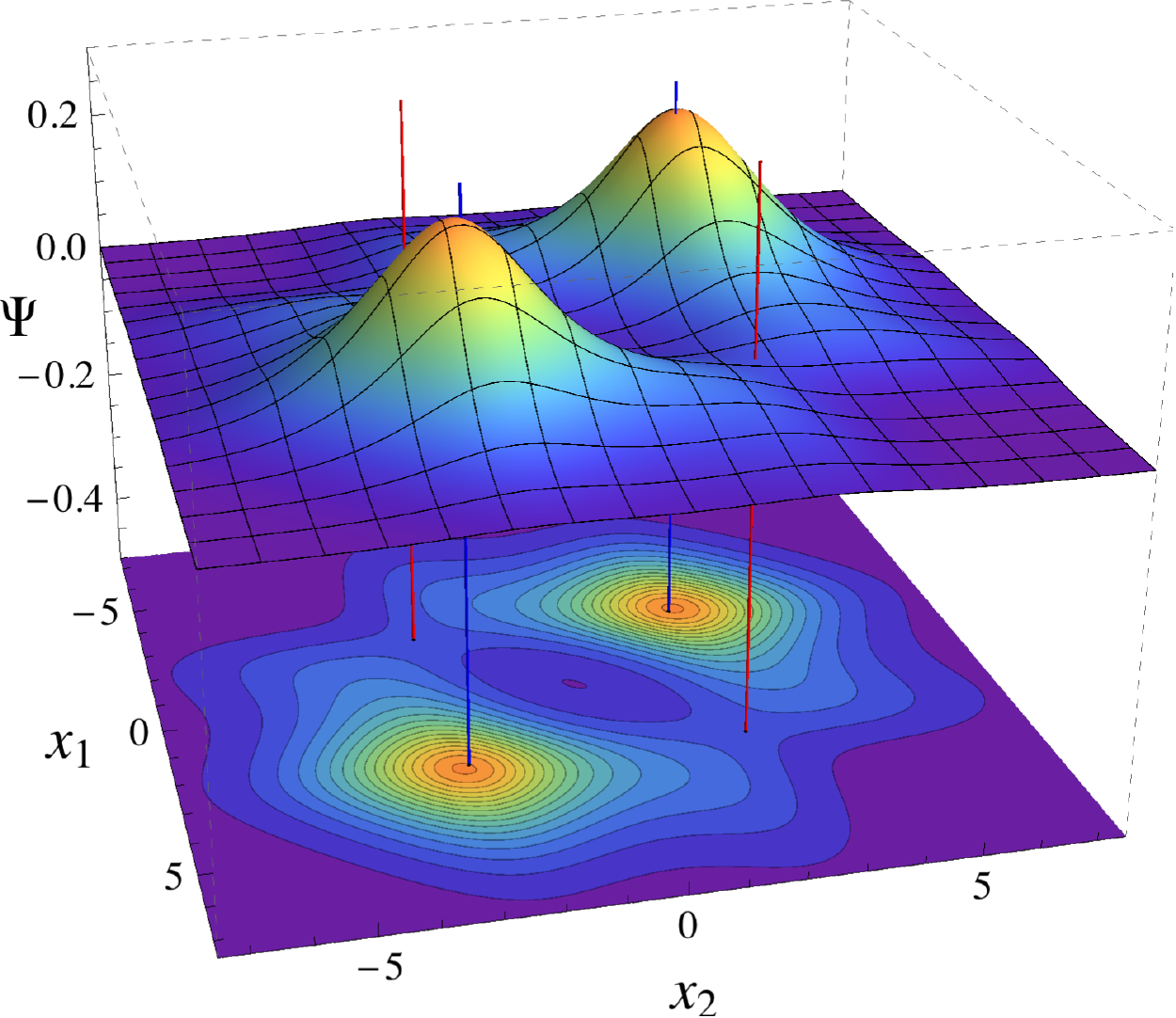}
  \caption{The same as in Fig.~\ref{fig:HFat5Bohr}, though the correlation function is now calculated from its differential equation (Eqs~\eqref{eq:fStat} and~\eqref{eq:f12NumDifEq}) with the HF orbital as input for the integrals in $\mu(x)$ and $\nu(x)$.}
  \label{fig:HFf12at5Bohr}
\end{figure}

Although we only considered a 1-dimensional hydrogen molecule, the HF error in the dissociation limit is still present. Due to the ionic contributions, the energy becomes way too high and behaves as $-1/(2R_{\text{H--H}})$, instead of becoming constant. The spurious ionic contributions can easily be made visible by plotting the HF wave function as has been done in Fig.~\ref{fig:HFat5Bohr} for $R_{\text{H--H}} = 5.0$ Bohr. In the upper panel we plot the normalized orbital, $\norm{\phi} = 1$, as a function of the distance from the bond mid-point and the correlation function, $f(x)$, which is normalized such that $\norm{\Psi} = 1$. In the lower panel the full HF wave function is shown. Vertical lines have been drawn at the nuclear positions. The red lines indicate the position when both electrons are on the same atom (ionic configuration) and the blue lines indicate the position when both electron reside on different atoms (covalent configuration). The HF wave function has equal peaks at all these configurations, so has equal covalent and ionic contributions. We know, however, that the atoms of a dissociated hydrogen molecule can be considered as independent atoms, so the real wave function of a H$_2$ molecule should only show peaks at the covalent (blue) positions as in Fig.~\ref{fig:exactWaveAt5Bohr}. Indeed, when we solve for the correlation function, the wave function is greatly improved as shown in Fig.~\ref{fig:HFf12at5Bohr}. The correlation function is able to squash down the ionic peaks of the HF wave function by having a small amplitude for short inter-electronic distances, hence reducing the contribution of the ionic configurations. Most of the amplitude is now on the covalent positions and only small ridges towards the ionic positions remain. The improved wave function also gives a much better energy (Fig.~\ref{fig:energies}). However, the ridges towards the ionic positions still give a significant contribution and cause the still rather large overshoot of the energy. At longer bond distances the energetic contribution of these ionic ridges is reduced and the total energy becomes closer to the exact one, which explains the appearance of the bump if only the correlation function is optimized without any relaxation of the HF orbital. The appearance of such a bump reminds a lot of the famous bump in the H$_2$ dissociation of the random-phase approximation (RPA) on top of a Kohn--Sham 
calculation~\cite{FuchsNiquetGonze2005, DahlenLeeuwenBarth2006}. Even if the Kohn--Sham orbitals are fully optimized, the RPA bump still persists~\cite{CarusoRohrHellgren2013}. The bump from our correlated Ansatz, however, completely disappears when the orbital and correlation are fully optimized together, as we will show in Sec.~\ref{sec:fullOpt}.

\begin{figure}[t]
  \includegraphics[width=0.5\columnwidth]{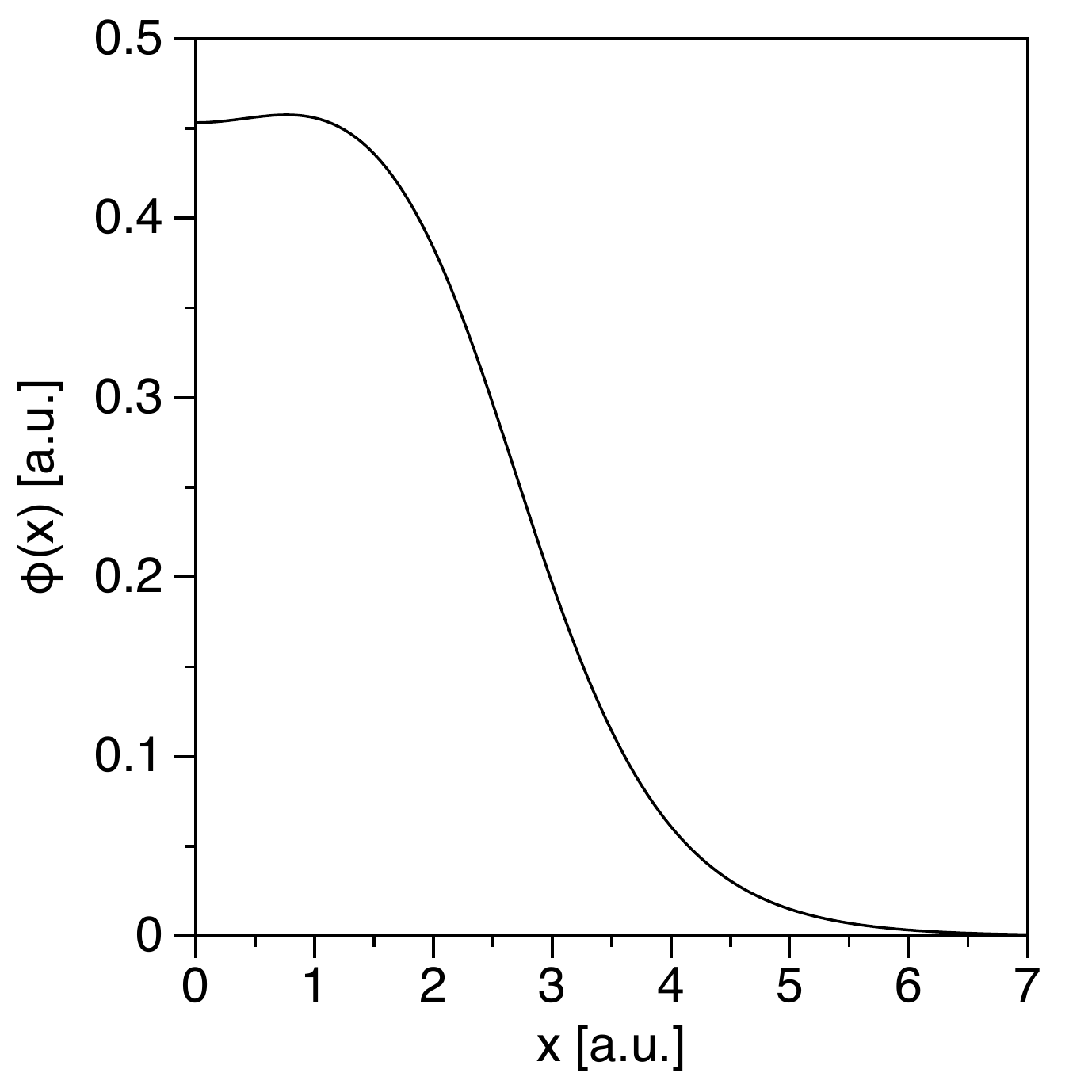}%
  \includegraphics[width=0.5\columnwidth]{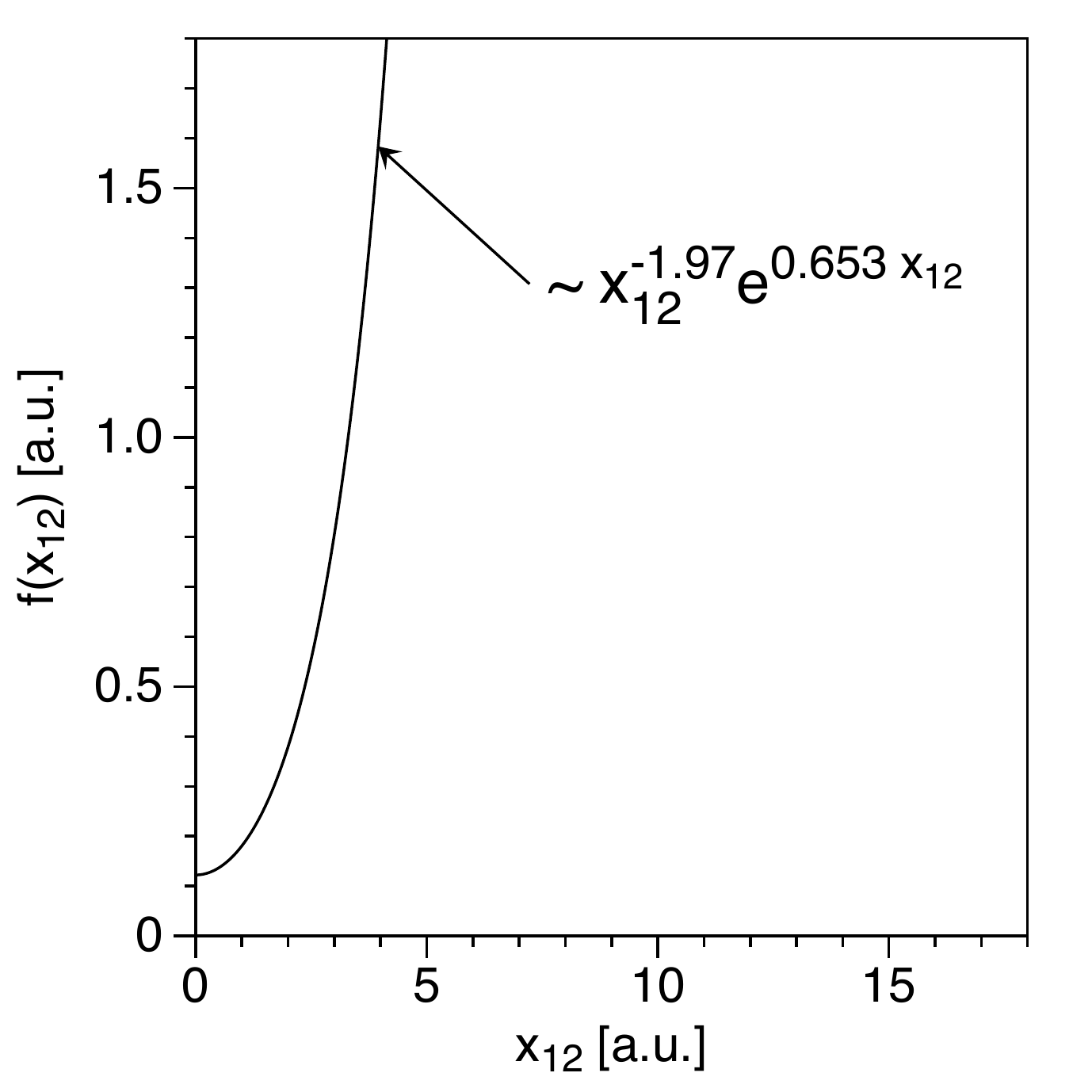}
  \includegraphics[width=\columnwidth]{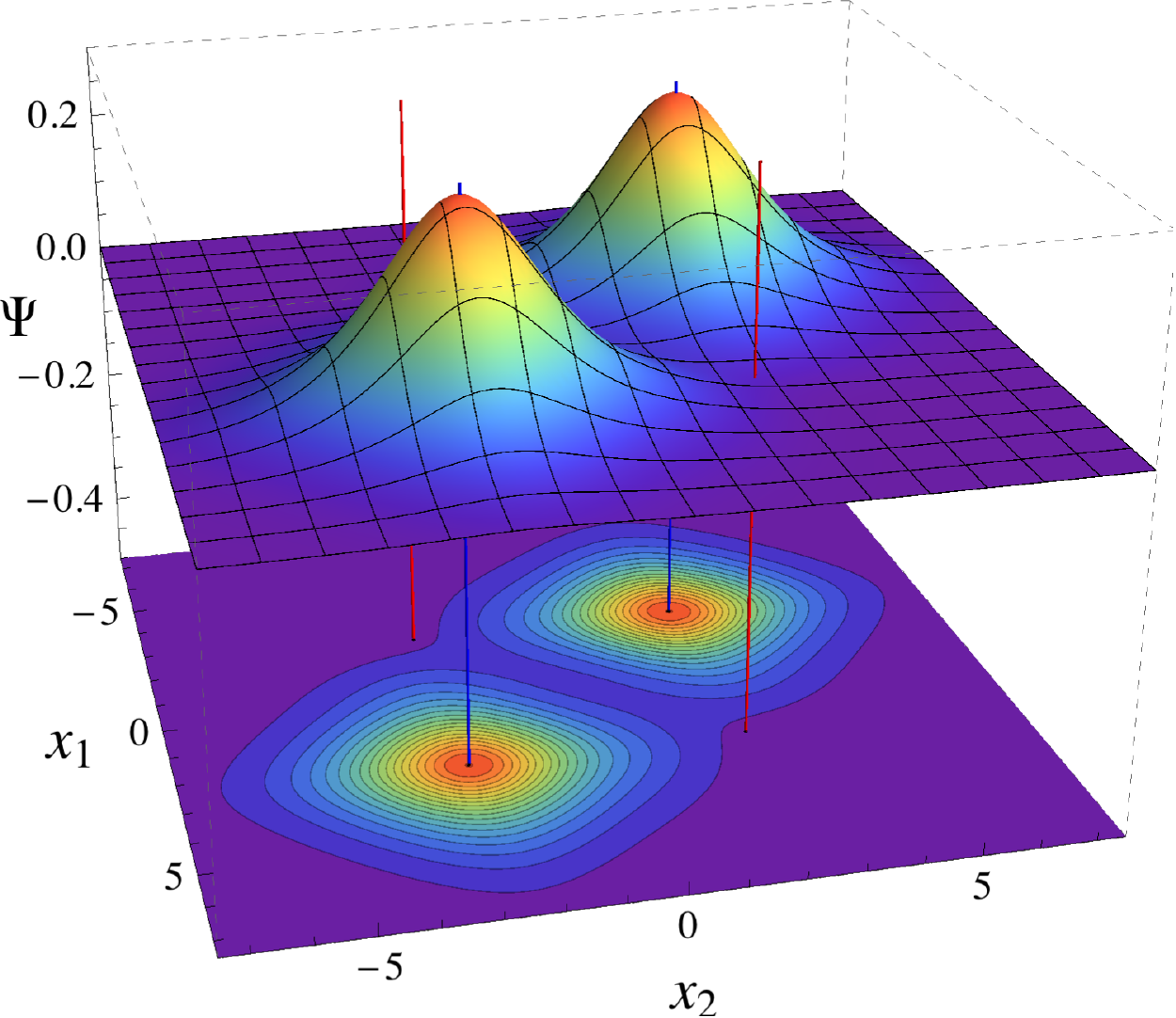}
  \caption{The same as in Fig.~\ref{fig:HFat5Bohr}, though the orbital and correlation function have now been fully optimized to give the minimal energy.}
  \label{fig:fullF12at5Bohr}
\end{figure}

\subsection{Full optimization of orbital and correlation function}
\label{sec:fullOpt}

\begin{figure}[t]
  \includegraphics[width=\columnwidth]{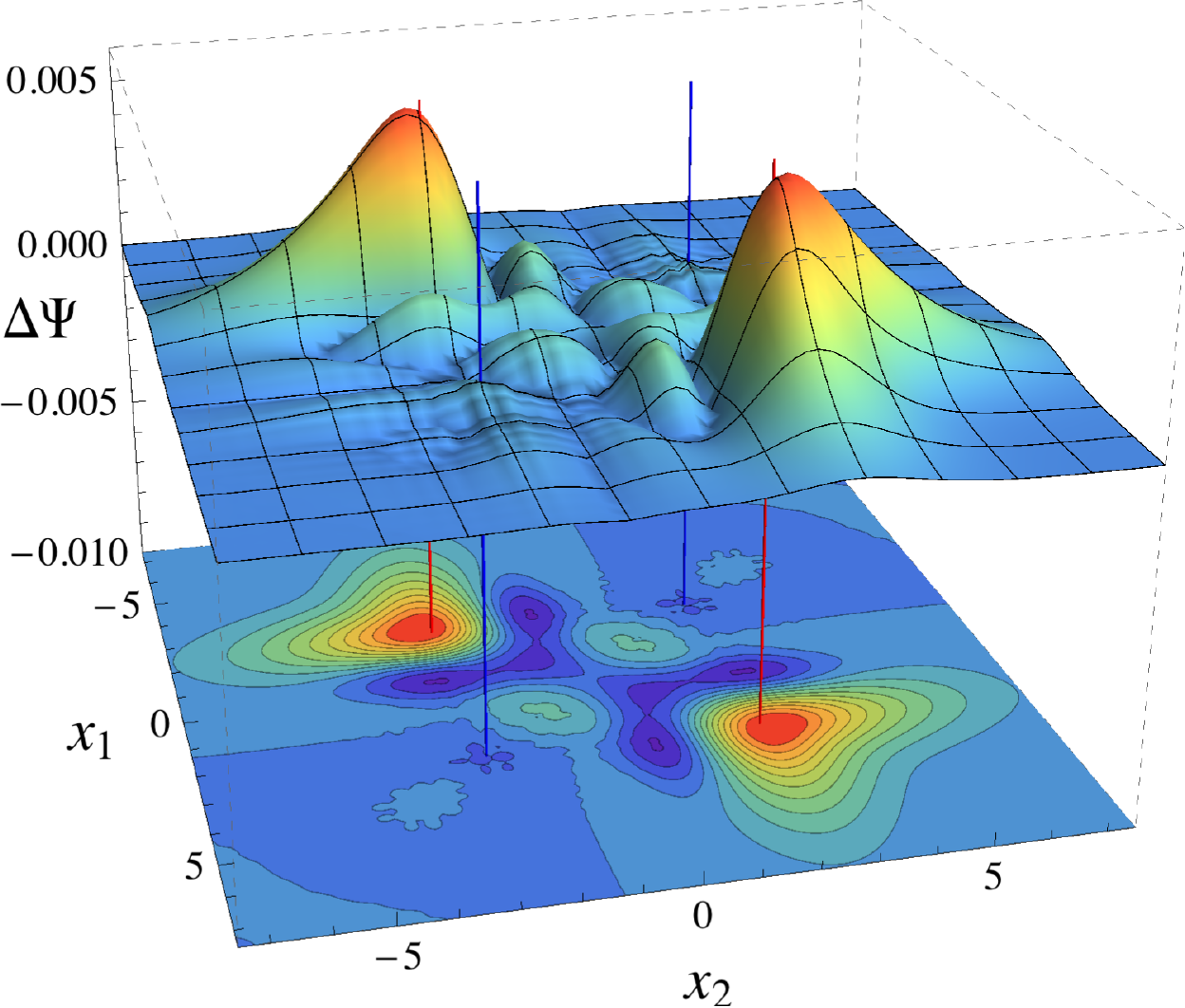}
  \caption{The difference between the exact wave function, $\Psi^{\text{exact}}$, and the fully optimized, explicitly correlated wave function~\eqref{eq:explCorWave1D}, $\Psi^{\phi^2\!f}$, defined as $\Delta\Psi \isDefinedAs \Psi^{\text{exact}} - \Psi^{\phi^2\!f}$.}
  \label{fig:diffWaveAt5Bohr}
\end{figure}

The results for the orbital, correlation function and the full wave function are shown in Fig.~\ref{fig:fullF12at5Bohr}. The ionic ridges in the partially optimized wave function (Fig.~\ref{fig:HFf12at5Bohr}) are a remnant of the ionic peaks of the HF wave function. By shifting the orbital towards the bond midpoint, the orbital relaxation is able to remove these ionic ridges from the fully optimized wave function. The lack of amplitude of the fully optimized orbital on the nuclei has now to be compensated for by a diverging correlation function (Fig.~\ref{fig:fullF12at5Bohr} upper right panel). The combination of these additional relaxations of both the orbital and the correlation function are so effective in describing the physics, that the full wave function becomes almost identical to the exact wave function on this scale, cf.\ Fig.~\ref{fig:exactWaveAt5Bohr}. The remaining difference between the two wave functions is shown in Fig.~\ref{fig:diffWaveAt5Bohr}. The main difference is a lack of amplitude of the explicitly correlated wave function at the ionic configurations. One could say that the simple Ansatz, $\Psi^{\phi^2\!f}$, is over-correlated, because the main error is a lack of ionic contributions which could in principle be included by adding a small fraction of the HF determinant. However, this difference is small on the overall scale of the wave function, so the total energy of the fully optimized wave function and the energy from the exact diagonalization are nearly indistinguishable in Fig.~\ref{fig:energies}. The maximum difference in energy occurs around $R_{\text{H--H}} \approx 5.0$ Bohr and is less than one mHartree.

An other interesting aspect are the positions of the maxima of the wave function at the different stages of the optimization. In the case of the HF wave function we find that the maxima are somewhat contracted towards the bond midpoint (Fig.~\ref{fig:HFat5Bohr} lower panel). This contraction is due to the ionic configurations of the wave function, since their contribution to the energy is lowered by moving the charge density on the negative ion, H$^-$, towards the positive ion, H$^+$. When we subsequently optimize the correlation function while keeping the HF orbital, the charge density is actually pushed outwards from the molecule (Fig.~\ref{fig:HFf12at5Bohr} lower panel). The correlation function pushes almost all the wave function amplitude away from the ionic configuration ($x_{12} \approx 0$). In doing so, however, part of the charge is also pushed outwards from the covalent contributions. A full relaxation of the explicitly correlated wave function allows the orbital to contract towards the bond midpoint to compensate for this spurious outward polarization. The maxima of the fully optimized, explicitly correlated wave function are now correctly localized at the nuclei (Fig.~\ref{fig:fullF12at5Bohr} lower panel).

\subsection{Bond distance dependence of correlation function and orbital}

\begin{figure*}[t]
  \includegraphics[width=\columnwidth]{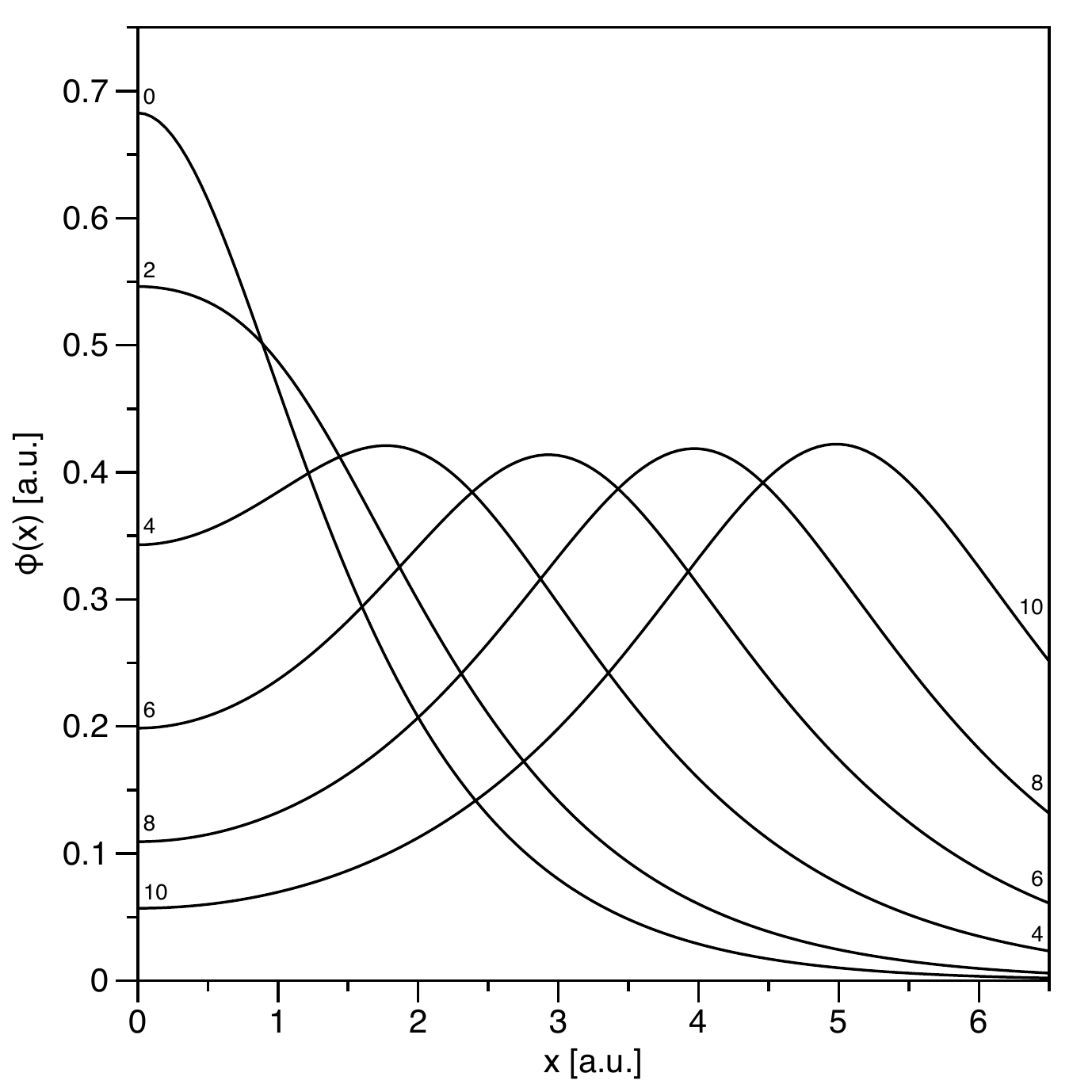}%
  \hfill%
  \includegraphics[width=\columnwidth]{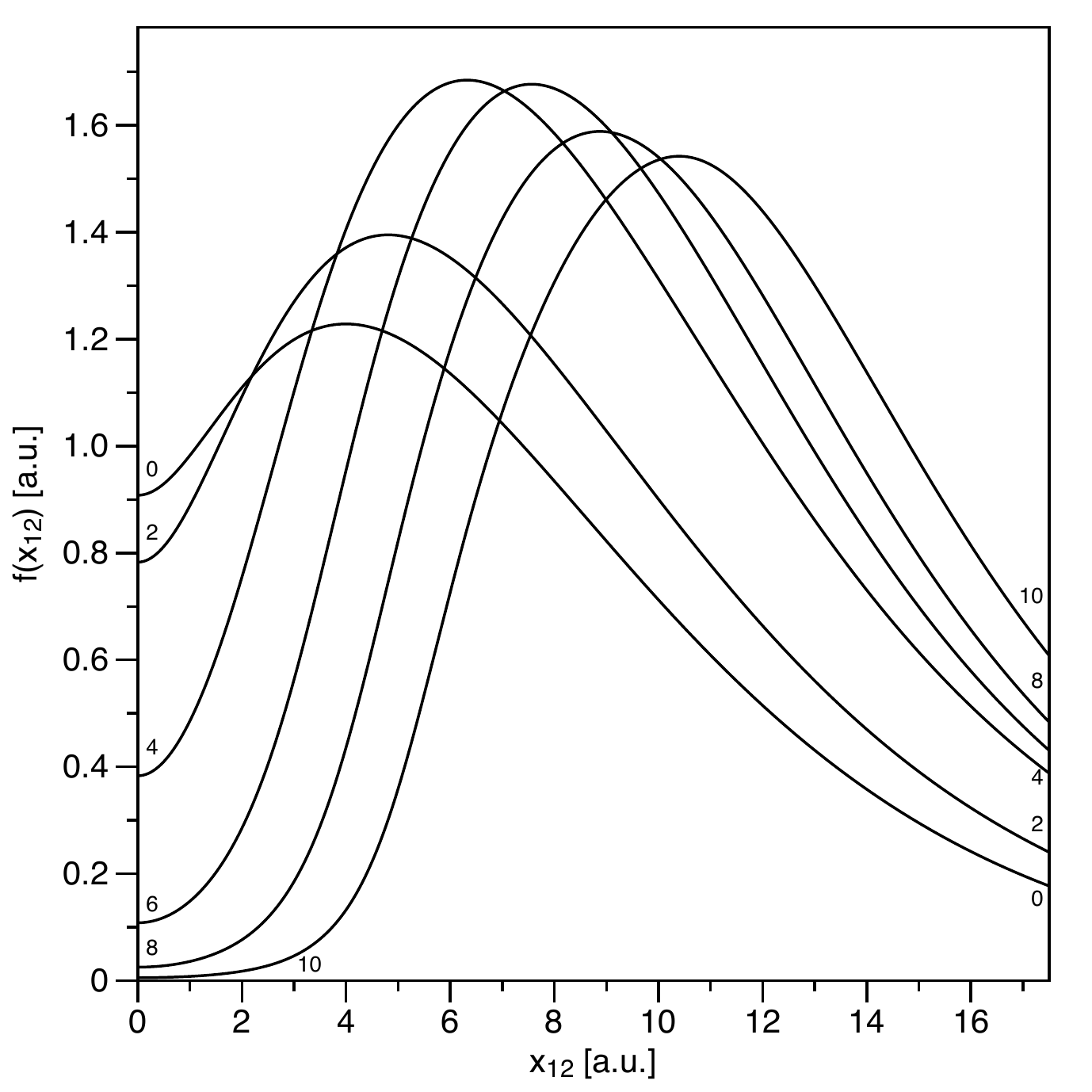}
  \begin{tabular}{@{}p{\columnwidth}@{\hspace{\columnsep}}p{\columnwidth}@{}}
    \caption{The HF orbitals for bond distances $R_{\text{H--H}} = 0.0, 2.0, \dotsc, 10.0$ Bohr. The bond distances are indicated by the small numbers in the figure.}
    \label{fig:HForbs} &
    \caption{The correlation functions optimized from the HF orbitals (Fig.~\ref{fig:HForbs}) for bond distances $R_{\text{H--H}} = 0.0, 2.0, \dotsc, 10.0$ Bohr. The bond distances are indicated by the small numbers in the figure.}
    \label{fig:HFf12s}
  \end{tabular}
  \includegraphics[width=\columnwidth]{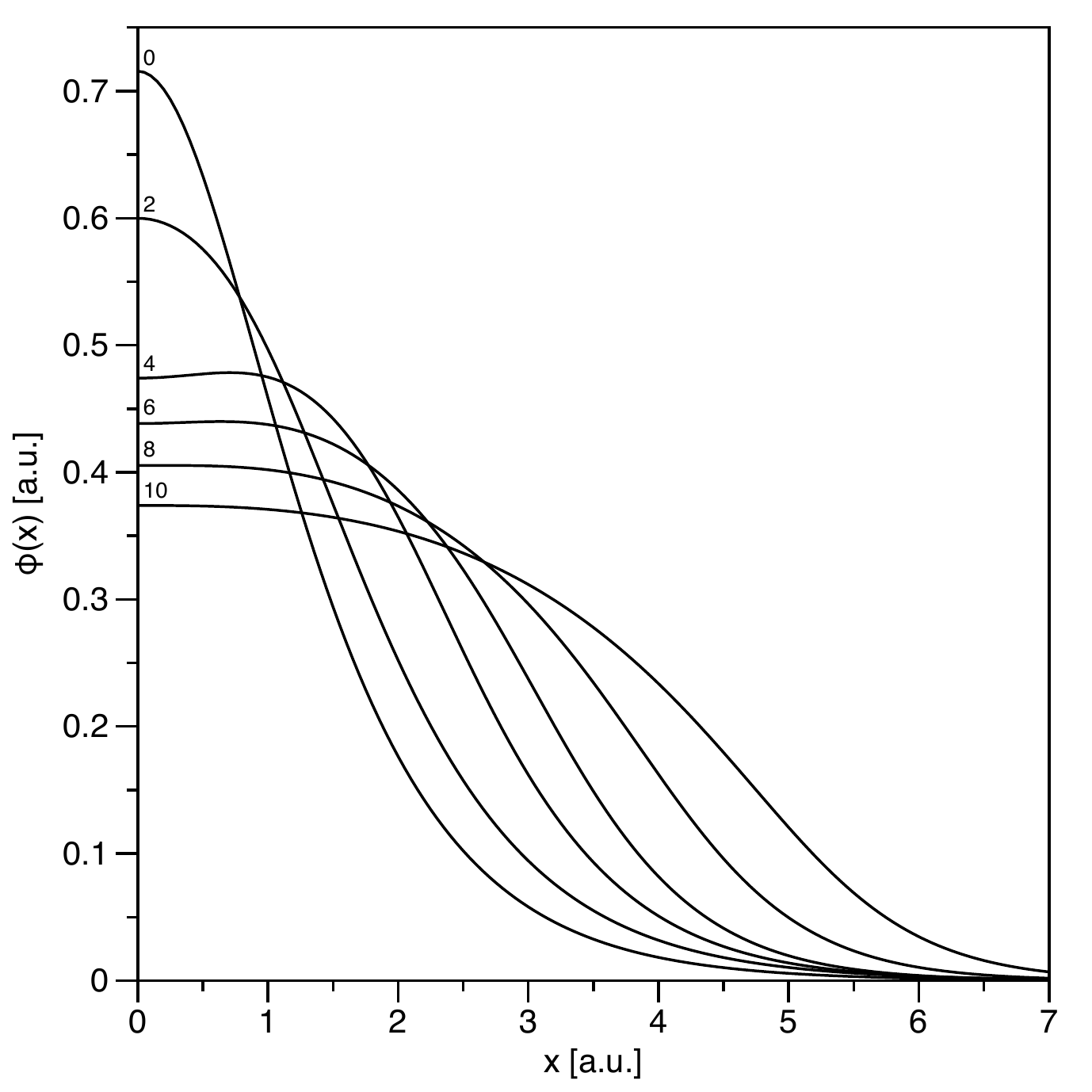}%
  \hfill%
  \includegraphics[width=\columnwidth]{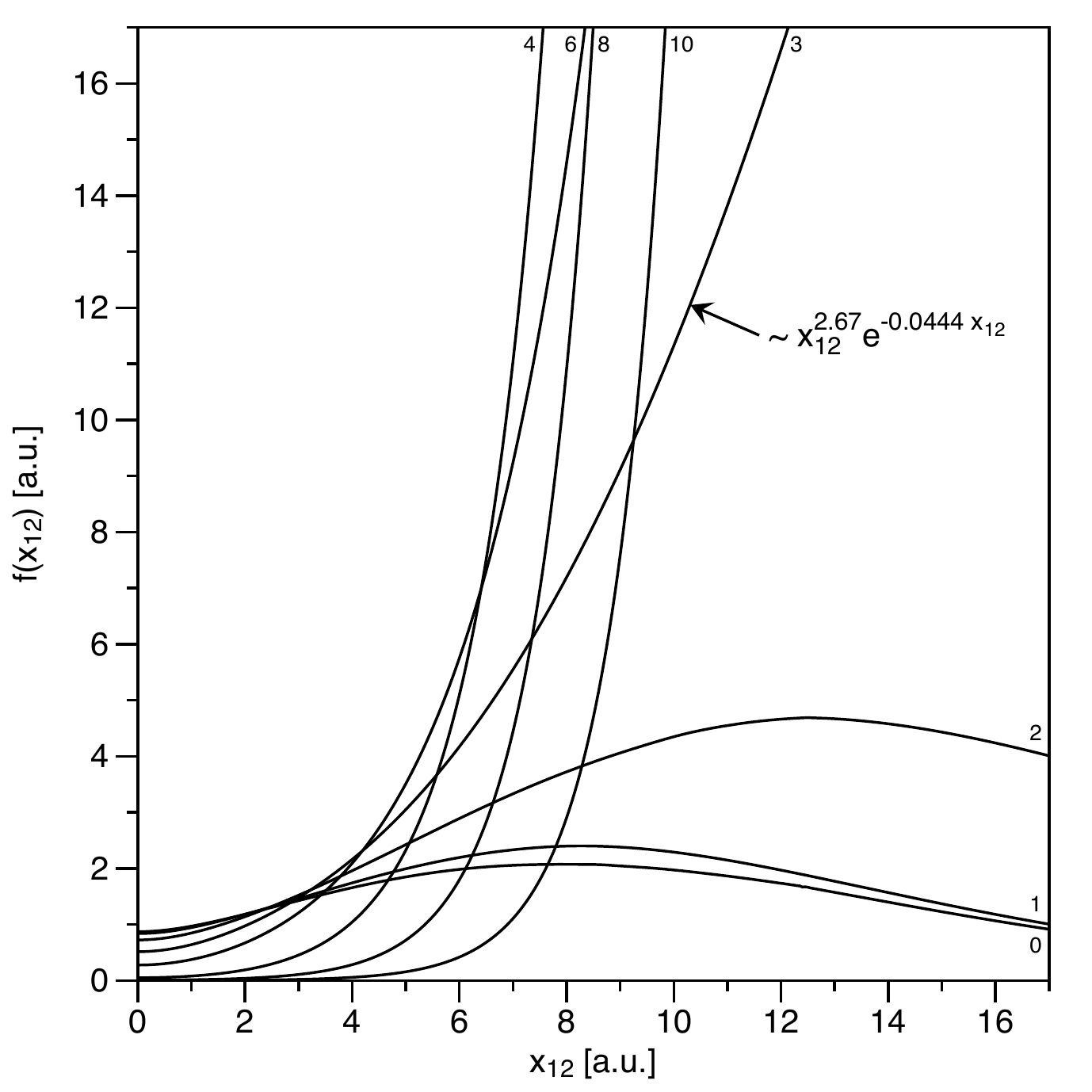}
  \begin{tabular}{@{}p{\columnwidth}@{\hspace{\columnsep}}p{\columnwidth}@{}}
    \caption{The fully optimized orbitals for bond distances $R_{\text{H--H}} = 0.0, 2.0, \dotsc, 10.0$ Bohr. The bond distances are indicated by the small numbers in the figure.}
    \label{fig:orbs} &
    \caption{The fully optimized correlation functions for bond distances $R_{\text{H--H}} = 0.0, 1.0, 2.0, 3.0, 4.0, 6.0, 8.0, 10.0$ Bohr, corresponding to the fully optimized orbitals in Fig.~\ref{fig:orbs}. The bond distances are indicated by the small numbers in the figure.}
    \label{fig:f12s}
  \end{tabular}
\end{figure*}

The situation described at a bonding distance of $R_{\text{H--H}} = 5.0$ Bohr is a good representative for the effects of the correlation function at the other bond distances, though the magnitude of the effects are different. The HF orbitals for bond distances $R_{\text{H--H}} = 0.0, 2.0, \dotsc, 10.0$ Bohr are shown in Fig.~\ref{fig:HForbs}. The HF orbital mainly tries to maximally decrease the potential energy from the nucleus-electron interaction, since this is the contribution that can reduce the energy most. Hence the HF orbital always has a maximum where the potential coming from the nuclei~\eqref{eq:nucPot} has a minimum. 
Since the full HF wave function is a simple product of the HF orbital, the HF wave function will always have a too large amplitude in regions where the electrons are near each other. In the unified atom limit (helium), this means that the HF wave function predicts a too high probability to find both electrons on the same side of the nucleus, i.e.\ in a polarized state. For a finite bond distance, the molecular case, this means that the HF wave function gives a too high probability to find both electrons at the same nucleus, i.e.\ to find the molecule in an ionic state. 
However, the energetic effect of these spurious ionic contributions is small for short bond distances and does not affect the energy too much (Fig.~\ref{fig:energies}). For larger bond distances the energetic error induced by the ionic configurations becomes more significant and cause the overshoot in the total energy.

If the correlation function is added to the wave function and optimized while retaining the HF orbital, the spurious ionic contributions are removed from the HF wave function. The correlation function achieves this by having a small value for small interelectronic distances ($x_{12} \approx 0$) and by having a large value in the outside region. This is the general feature of the correlation function at all bond distances as can be seen in Fig.~\ref{fig:HFf12s}. Although the correlation function is allowed to diverge for large $x_{12}$, it always decays for $x_{12} \to \infty$. The lack of divergence is understandable, since in the ionic contributions of the HF wave function the electrons are close to each other, so there is a strong repulsion. To reduce this repulsion, the HF orbital becomes more diffuse to provide the electrons ``additional space'' to avoid each other better. Since the HF orbitals are too diffuse, the HF density becomes too diffuse as well. A diverging correlation function would push the wave function even further away from the nuclei which would lead to a significant increase in energy. The correlation function solved from the HF orbital should therefore always be an asymptotically decaying function. Although the correlation function is quite effective at removing the ionic configurations from the HF wave function, small ionic ridges always remain due to the large amplitude of the HF orbital at the nuclei, as discussed before in the particular case of $R_{\text{H--H}} = 5.0$ Bohr. Though these ionic ridges become longer for more elongated bond distances, their contribution to the energy reduces (see Fig.~\ref{fig:energies}).

When the orbital and correlation are fully relaxed, the orbital contracts towards the bond midpoint to remove the ionic ridges (compare Fig.~\ref{fig:orbs} with Fig.~\ref{fig:HForbs}). The maximum of the orbital is typically not located anymore at the minima of the nuclear potential~\eqref{eq:nucPot}. Additionally the decay rate of the orbital is strongly increased for elongated bond distances, allowing for a diverging correlation function (See Fig.~\ref{fig:f12s}) while still retaining a sufficiently compact density near the nuclei.

However, for short bond distances the correlation function is still a decaying function, though the steepness of the correlation function grows for increasing bond lengths. Slightly beyond $R_{\text{H--H}} = 4.0$ Bohr the correlation function needs to become divergent to obtain the required steepness. The divergence is steadily increased until $R_{\text{H--H}} \approx 6.0$ Bohr where the exponent of the correlation function saturates at $\gamma_f \approx 1.0$. Elongating the bond even further only causes the correlation function to shift to the right and it remains equally divergent, i.e.\ the exponent remains approximately the same.
We can understand this behavior from the fact that in the dissociation limit the exact solutions becomes simply a symmetrized product of one electron located on the left nucleus and the other on the right nucleus. This is the well known Heitler--London Ansatz~\cite{HeitlerLondon1927} which can be expressed as
\begin{align*}
\Psi^{\text{HL}}(x_1,x_2) = \frac{1}{\sqrt{2}}\bigl(a(x_1)b(x_2) + a(x_2)b(x_1)\bigr),
\end{align*}
where $a(x)$ and $b(x)$ are solutions for one particle in the potential of the right and left nucleus respectively. If we now fix electron 2 on the left nucleus, we can neglect the last term in the Heitler--London wave function
\begin{align}\label{eq:HLsimple}
\Psi^{\text{HL}}(x_1,-\rho) \approx \frac{1}{\sqrt{2}}a(x_1)b(-\rho),
\end{align}
where $\rho \isDefinedAs R_{\text{H--H}}/2$. The function $a(x)$ will be a hydrogen like function for our 1D H$_2$ system, so first increase exponentially towards the nucleus as $\e^x$ and decay exponentially as $\e^{-x}$ when we move away from the molecule. For the explicitly correlated wave function we find
\begin{align*}
\Psi^{\phi^2\!f}(x_1,-\rho) = \phi(x_1)\phi(-\rho)f(x_1+\rho).
\end{align*}
From Fig.~\ref{fig:orbs} we see that in the dissociation limit a plateau starts to develop around the bond-midpoint before the orbital starts to decay exponentially to zero, so the correlation function will have to deform the orbital such that we recover the one-electron solution $a(x)$ located at the right atom~\eqref{eq:HLsimple}. Hence, the correlation function has to diverge to push the orbital outwards and to create the proper asymptotic decay, $\e^{x}$ on the left side of the nucleus. The correlation function should therefore diverge as $\e^x$ and indeed the numerical solution at $R_{\text{H--H}} = 10.0$ diverges as $\e^{\gamma_f x}$ with $\gamma_f \approx 0.96$. Further outward, the asymptotic decay of the orbital and the correlation function have to combine correctly to give the proper asymptotic decay on the right side of the nucleus, $\e^{-x}$. Hence, the orbital should decay as $\e^{-2x}$. The fully optimized orbital at $R_{\text{H--H}} = 10.0$ decays as $\e^{\gamma_{\phi}x}$ with $\gamma_{\phi} \approx -2.16$, which is quite close to the expected value of $-2$.

\section{A natural amplitude assessment}
\label{sec:occVanish}

\subsection{Nonvanishing of NO occupation numbers}

In this section we will address the question whether the natural occupation numbers of the explicitly correlated wave function can become zero. We will use our theorem that the explicitly correlated wave function can only have vanishing occupation numbers if and only if the Fourier transform of the correlation function vanishes on an open set (cf.~Sec.~\ref{sec:introduction} and Ref.~\cite{GiesbertzLeeuwen2013a}). Unfortunately, the correlation function becomes divergent from $R_{\text{H--H}} \approx 4.0$ Bohr onwards, impairing a straightforward numerical calculation of its Fourier transform. To deal with this difficulty, one has to eliminate its divergent part in some manner. A typical strategy is to subtract the divergent part and deal with that analytically. Unfortunately this strategy does not work here, since due to the fractional power, $x^s$, in the asymptotic part, we would introduce a new divergency at the origin ($s < 0$ typically for divergent correlation functions). Instead we would like to ``borrow'' some of the asymptotic decay of the orbital to make the correlation function square integrable ($\in L^2$).  To show the idea we first 
shift the integration variable in condition~\eqref{eq:zeroNOcond} to $x_{12}$ and split the integral in two pieces
\begin{align*}
0 &= \integ{x_{12}} f(x_{12}) \chi_i(x_1 - x_{12}) \\
&= \int\frac{\ud x_{12}}{1 + \e^{2\eta x_{12}}} f(x_{12}) \chi_i(x_1 - x_{12}) \\
&\eqspace {} + 
\int\frac{\ud x_{12}}{1 + \e^{-2\eta x_{12}}} f(x_{12}) \chi_i(x_1 - x_{12}).
\end{align*}
The factors $1/(1+\e^{\pm2\eta x_{12}})$ in the integrals effectively split the original integral in a left and right part. Since they are not identically each others negatives and their sum has to add up to zero for any separation $\eta$, they both need to be zero independently.
If we use
\begin{equation*}
\frac{2}{1+\e^{\pm2\eta x_{12}} } =  \frac{ \e^{\mp \eta x_{12}}}{\cosh (\eta x_{12})}
\end{equation*}
and multiply the integrals with $\e^{\pm \eta x_1}$ we can write the zero condition on
both integrals as
\begin{align*}
\integ{x_{12}} g(x_{12}) \e^{\pm\eta(x_1 - x_{12})}\chi_i(x_1 - x_{12}) = 0,
\end{align*}
where we defined
\begin{align*}
g(x_{12}) \isDefinedAs \frac{f(x_{12})}{\cosh(\eta x_{12})}.
\end{align*}
The cosine hyperbolic function is conveniently chosen to preserve the symmetry and eliminate the divergency on both sides.
after a suitable choice of $\eta$ to be discussed below.
Now we can proceed as before in the introduction~\cite{GiesbertzLeeuwen2013a} and take the Fourier transform to derive the condition
\begin{align*}
\Fourier[g](k) \cdot \Fourier\bigl[\e^{\pm\eta x}\chi_i(x)\bigr](k) = 0.
\end{align*}
Using the same arguments as before we find that vanishing occupation numbers only exist if and only if the Fourier transform of the regularized correlation function $\Fourier[g](k)$ vanishes on a finite interval.
The regularization parameter $\eta$ should be chosen sufficiently high to make the new function $g \in L^2$.  By this regularization of the correlation function we see that $\chi_i$ has obtained an additional divergent factor, so the regularized correlation function effectively ``borrows'' asymptotic decay from $\chi_i$. Since the NO in $\chi_i(x) \isDefinedAs \phi(x)\varphi_i(x)$ already puts $\chi_i \in L^2$, we can ``borrow'' the full exponential decay from the orbital, $\phi(x)$, and still have $\e^{\pm\eta x}\chi_i(x) \in L^2$. 
By choosing $\eta = \gamma_{\phi}$ we see (see also the analysis in the Appendix~\ref{ap:technics}) that $g(x)$ is square integrable  and also $\e^{\pm \eta x} \chi_i (x)$ since
the natural orbitals $\varphi_i (x)$ are exponentially decaying and $\e^{\pm \eta x} \phi (x)$ grows maximally as $x^{p_{\phi}}$.

\begin{figure}[t]
  \includegraphics[width=\columnwidth]{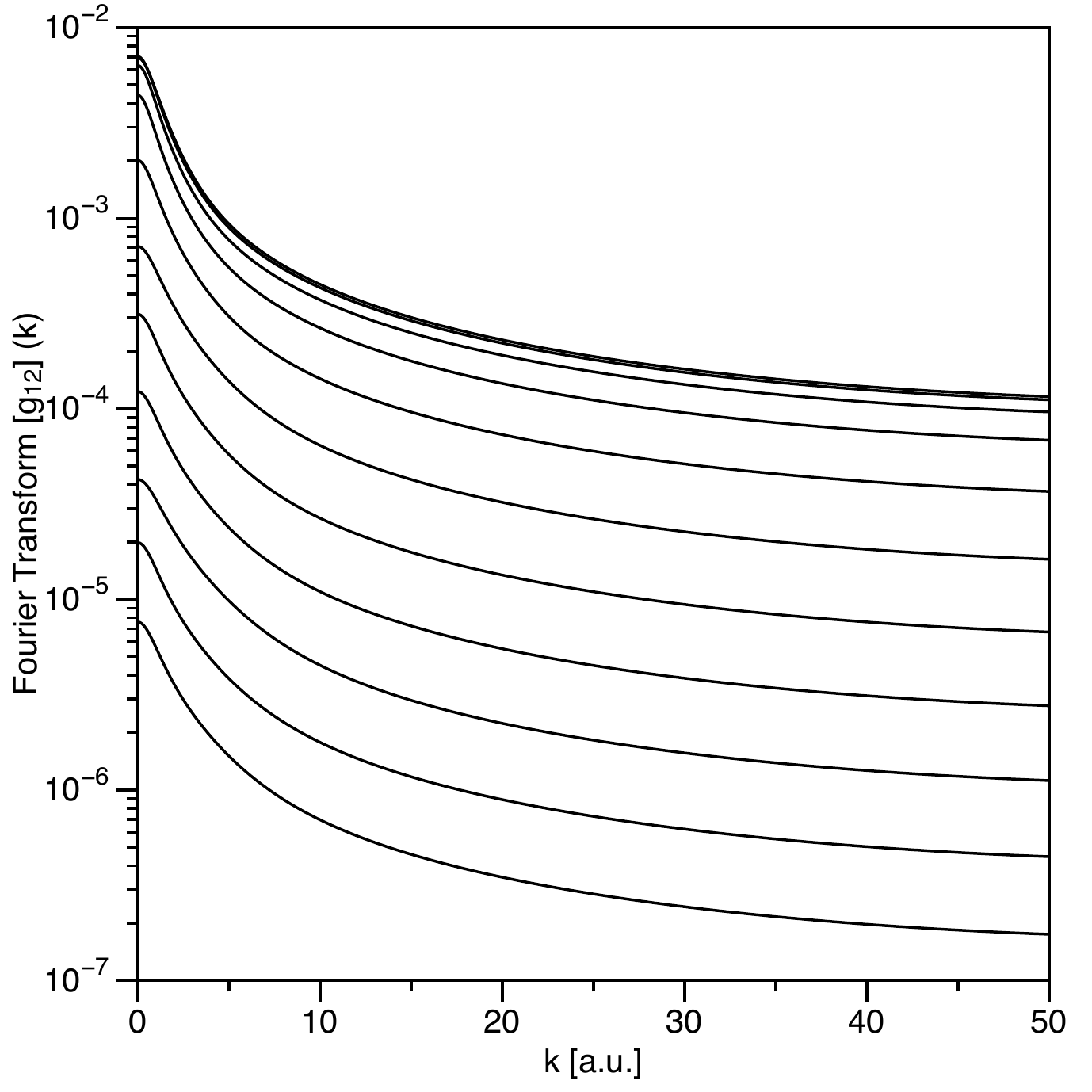}
  \caption{The Fourier transform of the regularized correlation function, $g(x_{12})$, for bond distances $R_{\text{H--H}} = 0.0, 1.0, \dotsc, 10.0$ Bohr (top to bottom) and $\eta = \gamma_{\phi}$. The curves for $R_{\text{H--H}} = 0.0$ Bohr and $R_{\text{H--H}} = 1.0$ Bohr are very close to each other, so they seem to form one thick line together.}
  \label{fig:g12Four}
\end{figure}

Since we showed that the Fourier transform of $g(x_{12})$ can be used equally well as the Fourier transform of $f(x_{12})$, we can use this result to show that our fully optimized explicitly correlated wave function does not have any vanishing occupation number. After regularizing the correlation function, we have calculated the Fourier transform numerically for bond distances of $R_{\text{H--H}} = 0.0, 1.0, \dotsc, 10.0$ Bohr. The Fourier transform of the regularized correlation functions are plotted in Fig.~\ref{fig:g12Four}. The Fourier transforms of $g(x_{12})$ are smooth functions and do not vanish in the plotted region. The maximum $k$-value included in the plot ($k=50$) corresponds to a discretization of $\Delta x = \pi/50 \approx 0.063$ Bohr. We have not observed any structure smaller than 0.063 Bohr in the correlation function, so the Fourier transform of $g(x_{12})$ should be analytic for larger values of $k$ as well. Hence, $\Fourier[g](k)$ can neither vanish outside the plotted region on a finite region, so the natural occupation numbers cannot become zero for our calculated wave functions. Assuming a smooth behavior of the regularized correlation function between the calculated distances, we reach the same conclusion over the whole range $R_{\text{H--H}} = [0.0,10.0]$ Bohr: the natural occupation numbers do not vanish. Since the Fourier transform of the regularized correlation function is only going down without changing its shape too much (see Fig.~\ref{fig:g12Four}), we expect this statement to hold for any finite bond length beyond $R_{\text{H--H}} = 10.0$ Bohr.
We found a similar statement to be true for the correlation function of the partially  optimized wave function (in which case $f$ does not need to be regularized).

\subsection{Natural amplitudes and avoided crossings}

\begin{figure}[t]
  \includegraphics[width=\columnwidth]{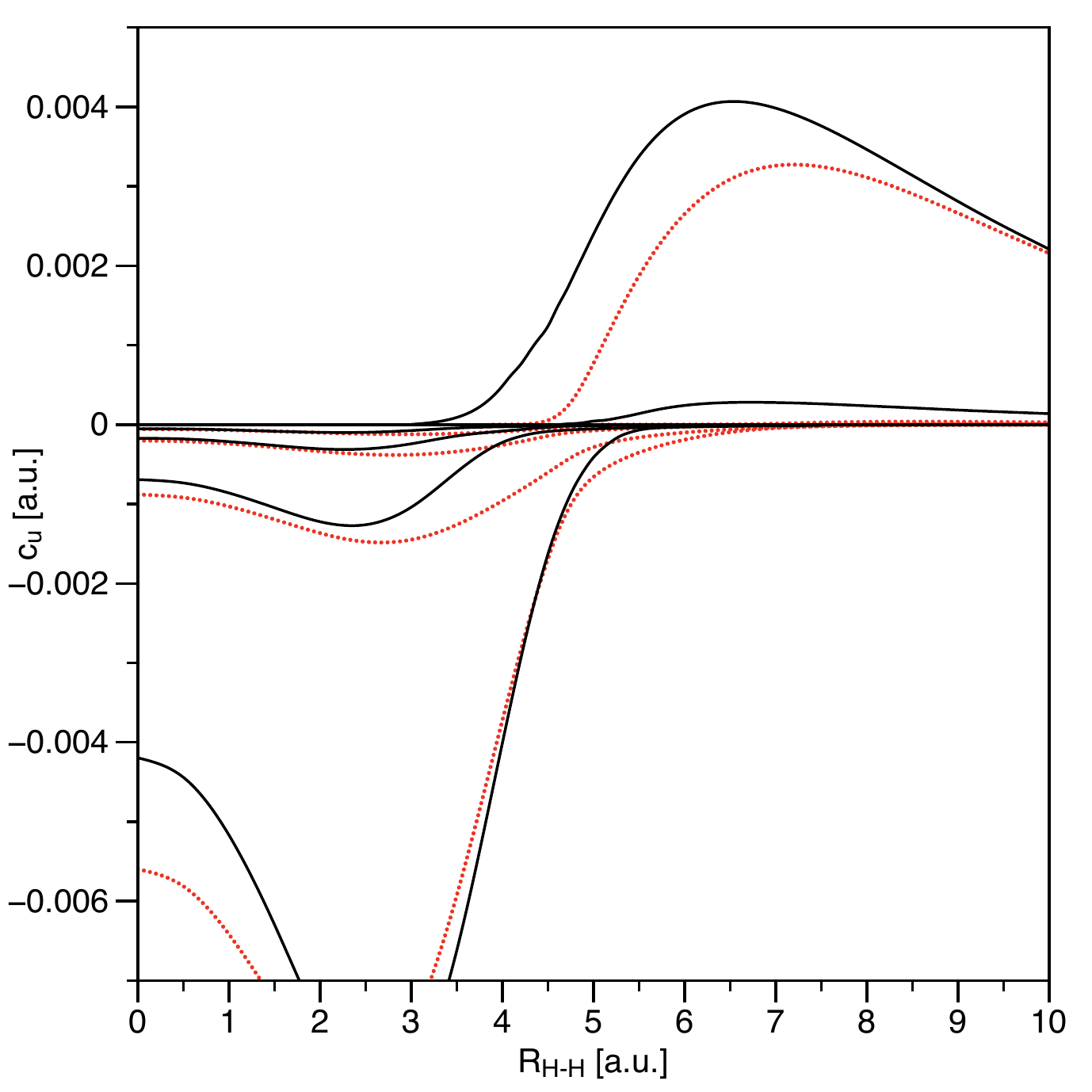}
  \caption{The ungerade NO amplitudes of the fully optimized Ansatz versus the bond distance (straight, black lines). As a reference, the numerically exact NO amplitudes are shown as red, dotted lines. The coefficient of the highest occupied ungerade NO are off the scale, so not visible.}
  \label{fig:coefF12}
\end{figure}

Though we have only demonstrated for the explicitly correlated approximation that unoccupied NOs do not exist, this does not necessarily imply that no vanishing natural occupation numbers are present in the exact case. From comparison with the numerically exact solution (Fig.~\ref{fig:diffWaveAt5Bohr}), however, we see that our approximate wave function provides a faithful representation of the exact wave function. 
Indeed, when we compare the numerical exact NO amplitudes with the NO amplitudes of the simple Ansatz (Fig.~\ref{fig:coefF12}), we observe that the behavior of the NO amplitudes is quite similar in both wave functions. The correlated orbital product correctly captures the avoided crossings of the expansion coefficients around $R_{\text{H--H}} \approx 5$ Bohr, which are due to the transition from a chemical bond to Van der Waals bond~\cite{CioslowskiPernal2006, ShengMentelGritsenko2013, GiesbertzLeeuwen2013b}. It is striking that the NO amplitudes of the simple Ansatz are more positive than the NO amplitudes of the exact wave function. We can understand this shift from the difference between the wave functions plotted in Fig.~\ref{fig:diffWaveAt5Bohr} for $R_{\text{H--H}} = 5.0$ Bohr. We see that the main remaining error in the correlated orbital product is actually a lack of ionic configurations, so this approximate wave function can be considered to be somewhat over-correlated. We have shown before that the positive NO amplitudes are signature of the long-range Van der Waals effects~\cite{GiesbertzLeeuwen2013b}. Since $\Psi^{\phi^2\!f}$ is somewhat over-correlated, the Van der Waals effects will be overemphasized compared to the exact wave function to lower the total energy, which is reflected in the enhanced positive NO amplitudes. 

The close resemblance of the NO amplitudes between the two wave functions shows that our Ansatz captures the behavior of the NO amplitudes very well. Since $\Psi^{\phi^2\!f}$ gives such an accurate representation of the numerically exact wave function, the exact wave function should also not have any vanishing natural occupation number. Hence, the result that our approximate wave function does not have any unoccupied NO, provides a reasonable argument that the natural occupations in the 1D hydrogen molecule do not vanish (in addition to the even stronger argument presented in Ref.~\cite{GiesbertzLeeuwen2013b}).

\begin{figure}[t]
  \includegraphics[width=\columnwidth]{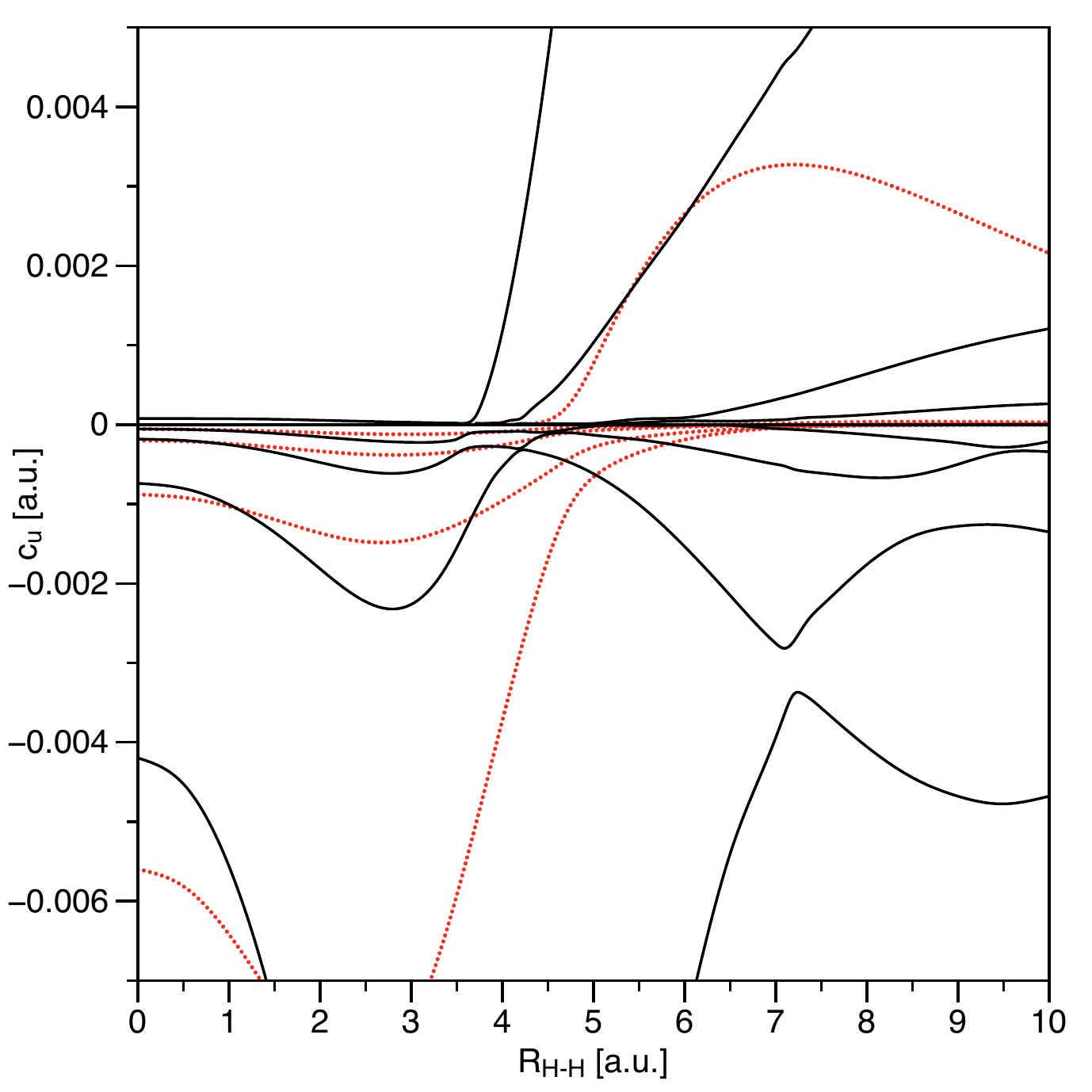}
  \caption{The ungerade NO amplitudes of the partially optimized Ansatz versus the bond distance (straight, black lines). As a reference, the numerically exact NO amplitudes are shown as red, dotted lines. The coefficient of the highest occupied ungerade NO are off the scale, so not visible.}
  \label{fig:coefHFf12}
\end{figure}

In Fig.~\ref{fig:coefHFf12} we show the NO amplitudes for the partially optimized Ansatz (HF orbital with optimized correlation function). At short bond distances the coefficients of the partially optimized Ansatz compare quite well with the exact amplitudes, since the HF orbital does not differ too much from the fully optimized one (see Fig.~\ref{fig:HForbs} and Fig.~\ref{fig:orbs}). Also the correlation functions do not differ too much (Figs.~\ref{fig:HFf12s} and~\ref{fig:f12s}), so the partially optimized Ansatz provides a good approximation to the exact wave function (since the fully optimized one always is). When the bond is stretched, however, the HF orbital starts to deviate strongly from the fully optimized one. The correlation function introduces a large amount of artificial correlation (too large amount of significant NO amplitudes) in its attempt to compensate for the bad HF Ansatz. The full optimization of the correlated orbital product relieves this difficulty and the quality of the approximate wave function is greatly improved, as is apparent from the NO amplitudes shown in Fig.~\ref{fig:coefF12}.

\section{Conclusion and outlook}
\label{sec:conclusion}

We showed that for a one-dimensional model of a two-electron diatomic molecule it is possible to construct a compact wave function that incorporates short range, Van der Waals and static near-degeneracy correlations.
This is achieved by optimizing a correlated orbital product Ansatz involving a correlation function $f(r_{12})$. Although the short range properties of the
correlation function are well known and fixed by the cusp condition, very little is known about the properties such a function needs to have in order to describe bond breaking as well 
as long-range Van der Waals type interactions. We elucidated the latter features. In particular we found that the correlation function needs to diverge for large $r_{12}$ at large 
internuclear distances while for shorter bond distances it has a characteristic maximum after which it decays. A natural amplitude analysis revealed that the restricted Ansatz gives
an accurate description of all three correlation effects, the short range, longe range and static correlations. The short range correlations induce non-vanishing NO amplitudes
whereas the pattern of avoided crossings shows that the long-range Van der Waals correlations are described well.  Static correlations lead to near degeneracy of the highest occupied
gerade and ungerade NO.

Although the divergent correlation function yields the correct features of the exact wave function it is undesirable to handle systems with more than 2 particles. Explicitly correlated wave functions typically only employ one correlation function for all electron pairs, e.g.
\begin{align*}
\Psi(\vecx_1,\dotsc,\vecx_N) = \sum_{i < j}f(r_{ij})\Phi(\vecx_1,\dotsc,\vecx_N),
\end{align*}
where $\Phi$ is an orbital based Ansatz again. If $\Phi$ would be only a single Slater determinant and such a wave function should describe a dissociating molecule, only the correlation function belonging to the dissociating electron pair should diverge and not for the other electron pairs. Providing each electron pair its own correlation function would not be feasible, since one would typically need to deal with $N(N-1)/2$ correlation functions. The only feasible way forward is to incorporate the static correlation effects in the orbital product Ansatz~\cite{FilippiUmrigar1996, VarganovMartineza2010, ZenCocciaLuo2014}.
For example, a wave function of the form
\begin{equation}
\Psi (\vecr_1,\vecr_2) = \bigl( \phi_1 (\vecr_1)   \phi_2 (\vecr_2)   +  \phi_1 (\vecr_2)   \phi_2 (\vecr_1) \bigr)  f(\vecr_{12} )  \nonumber
\end{equation}
in which $\phi_1$ and $\phi_2$ are general non-orthogonal orbitals will dissociate the H$_2$ molecule correctly even if $f$ would equal to one.
This means that less demands have to be put on $f$ which may prevent divergent behavior of the correlation function. This remains a future point of
investigation. Such an Ansatz can be further generalized to many-electron molecules. For instance, we can take the function $\Phi$ in Eq.~\eqref{eq:PsiF12Phi} to
be an anti-symmetrized product of non-orthogonal orbitals. This amount to the so-called Extended Hartree-Fock (EHF) Ansatz for $\Phi$~\cite{Lowdin1955a, Lowdin1955c, DahlenLeeuwen2001}.
If we then optimize both $\Phi$ and $f$ we obtain $N+1$ orbital equations of the form
\begin{align*}
\hat{H}_{\phi_j} [\left\{ \phi_k \right\} ,f] \phi_j &= E \, \phi_j \qquad \text{for $j = 1,\dotsc, N$}, \\    
\hat{H}_{f} [\left\{ \phi_k \right\} ] f &= E \, f,
\end{align*}
where $\hat{H}_{\phi_j}$ and $\hat{H}_{f}$ are effective Hamiltonians.
If we take $f=1$ then the first set of $N$ orbital equations reduce exactly to the EHF equations.
One practical strategy to incorporate the short range and Van der Waals correlations would be to first solve the EHF equations (some efficient ways of doing this have been
devised~\cite{VerbeekLenthe1991a, VerbeekLenthe1991b, ByrmanLentheVerbeek1993, PhD-Rashid2013}) and subsequently solve for the correlation function in the last equation.
This could provide a useful alternative for the F12 methods~\cite{Kutzelnigg1985, TewKlopper2005, KlopperManbyTen-No2006, KongBischoffValeev2011,HaunschildChengMukherjee2013} in which F12 has a fixed form rather than being determined by the shape of the molecular orbitals.
This approach is part of work in progress.

\begin{acknowledgments}
The authors acknowledge the Academy of Finland for research funding under Grant No.\ 127739. KJHG also gratefully acknowledges a VENI grant by the Netherlands Foundation for Research NWO (722.012.013).
\end{acknowledgments}

\appendix

\section{Numerical solution}
\label{ap:technics}

First we should know in what function spaces we should search for the optimal orbital and correlation function. If we consider the explicitly correlated ansatz~\eqref{eq:explCorWave1D} for $x_2 = x_1$, we see that the orbital should decay fast enough for the wave function to be integrable. Therefore, we require the orbital to be in the Sobolev space $H^1(\Reals)$, which is defined as
\begin{align}
H^1(\Reals) = \biggl\{f : \integ{x} \abs{f(x)}^2 + \abs{f'(x)}^2 < \infty\biggr\},
\end{align}
so the Sobolev space $H^1(\Reals)$ is equivalent to $L^2(\Reals)$ with the additional restriction that also the first order derivatives are also in $L^2(\Reals)$. So the Sobolev space not only guarantees that $\phi$ is square integrable but also smooth, which is necessary to give meaning to the Laplace operator in the variational equation.

Unfortunately we cannot make such statements for the correlation function $f$. Since the orbitals already decay, the correlation function is allowed to diverge, though, not too fast. Mathematically this can be made more precise by searching the correlation function in a weighted $H^1(\Reals)$-space defined as
\begin{align*}
H^1(\Reals,w) \coloneqq \left\{f : \integ{x}w(x)\bigl(\abs{f(x)}^2 + \abs{f'(x)}^2\bigr) < \infty \right\},
\end{align*}
where $w$ is called the weight. Note that for $w = 1$ we simply have $H^1(\Reals,1) = H^1(\Reals)$. Using this definition, we require the correlation function $f \in H^1(\Reals, w_{\phi})$, where $w_{\phi}$ is an orbital dependent weight
\begin{align*}
w_{\phi}[\phi](x) 
\coloneqq \integ{y}\Bgnorm{\phi\Bigl(\frac{x+y}{2}\Bigr)}^2\Bgnorm{\phi\Bigl(\frac{x-y}{2}\Bigr)}^2,
\end{align*}
implied by the normalization condition on the full wave function $\Phi$. Since the weight depends on the orbital, it seems hard to use this information in practice. However, we will see later when we work out our equations on a grid that this weight actually occurs quite naturally.

Especially since the function-space for the correlation function requires a lot of freedom, it is most natural to perform the calculation on a grid. Note that to evaluate the integrals~\eqref{eq:statIntegrals} we also need to interpolate the functions to off-grid values. Therefore, it is convenient to use cubic B-splines~\cite{PhD-Becke1981} for the calculation of the integrals and the derivatives, since they also provide immediately an interpolation scheme.

Since the full wave function is symmetric in $x_1$ and $x_2$, also the orbital and the correlation function are symmetric ($\phi(x) = \phi(-x)$ and $f(x) = f(-x)$), so we only need a grid to run from zero to infinity. The region around the molecule is most important and therefore, we like to have a good precision in this region. To provide good control, we divide the grid in two parts: the inner region around the molecule has a linear grid, $x(t) = t$ for $t \in [0,b]$, and the outer region is parametrized as
\begin{align*}
x(t) = t + \frac{(t - b)^3}{c - t}
\end{align*}
for $t \in (b,c)$, which ensures a sufficiently smooth transition between the inner and outer region. For the orbital we have used $b = \rho + 15.0$ and $c = b + 2.5$ with a density of 20 grid points per Bohr. We have included a large part of the asymptotic region, since the numerical integration is done on this grid and the correlation function is still quite sensitive to the precision of the numerical integration far away from the system. For the correlation function we have used $b = 25.0$ and $c = 27.5$ with a total of 550 grid points for all molecular distances.

Unfortunately, the use of splines does not lead a positive symmetric discretized version of the Laplacian. Therefore, we use the same trick as Becke~\cite{PhD-Becke1981} and write the Laplace operator formally as
\begin{align}
-\nabla^2 = \leftnabla \cdot \rightnabla,
\end{align}
where we used that the boundary terms vanish for $H^1$ functions. Using the discretization for the gradient, we obtain a positive symmetric discretized version of the Laplacian by construction. An additional advantage of this procedure is that we only have to consider the first order derivatives of the splines.

The equations are non-linear in character. Solving the stationarity equations~\eqref{eq:stationary} in a straightforward iterative manner is therefore not expected to converge. A more elegant approach is to use non-linear optimization techniques on the energy expression directly, together with its gradient~\eqref{eq:Egradient} to search for a minimum. We implemented a trust-region algorithm in combination with a symmetric rank one (SR1) approximation to the Hessian~\cite{NocedalWright1999, ConnGouldToint2000}. The downside of a direct energy optimization is that the energy is relative insensitive to the asymptotic behavior of the wave function. Since both the orbital and correlation function satisfy a differential equation, we can deduce that they will behave asymptotically as
\begin{align}\label{eq:asympBehavior}
\sim \left(q_0 + \frac{q_1}{x} + \frac{q_2}{x^2} + \dotsb\right)x^p\e^{-\gamma x}.
\end{align}
For the inner part of the orbital and correlation function we used their values at the grid points to specify their shape till a certain cut-off point, from whereon we used the parameters of the asymptotic expansion. Including the polynomial up to second order in $1/x$ gave good results. We used as cut-off parameters $\rho + 10.0$ for the orbital and 12.5 for the correlation function. When stretching the bond distance the cut-off parameter for the orbital was too ambitious for the full optimization, since the orbital is contracted towards the bond-midpoint. So we had to reduce the cut-off from $R_{\text{H--H}} = 7.0$ Bohr onwards. For the distances $R_{\text{H--H}} \in [7.0,8.0)$ Bohr we used $\rho + 7.5$ Bohr and for the distances $R_{\text{H--H}} \in [8.0,10.0]$ Bohr we used $\rho + 5.0$ Bohr as cut-off parameter for the orbital.

The complete optimization has been divided in three steps. First we set the correlation function to a constant ($f = 1$) and only optimize the orbital, i.e.\ the Hartree--Fock solution. Once we have found the HF orbital, we calculate the corresponding correlation function directly from its differential equation~\eqref{eq:fStat}. As mentioned in Sec.~\ref{sec:theory}, since the operator only depends on the orbital, this can be straightforwardly done by diagonalization. Direct calculation of $\mu(x)$ and $\nu(x)$ becomes numerically unstable in the asymptotic region, due to the division of two small inaccurate numbers. Therefore, the asymptotic behavior of these quantities has to be evaluated analytically. The analysis is rather straightforward, though technical, so it has been deferred to Appendix~\ref{ap:asympMuAndNu}.

There are two additional complications due to the discretization which need to be taken into account. The first one is that $f \notin H^1(\Reals)$, so the boundary conditions used in the discretization of the derivatives do not apply. However, $\sqrt{w_{\phi}(x)}f(x) \in H^1(\Reals)$, so we should solve for this function instead of the correlation function directly. The complication is that the friction term ($\mu(x)\du_x$ spoils the symmetry of the discretized operator. We can use a similar trick as with the Laplacian to make this part of the operator symmetric as well. Taking both considerations into account, the actual differential equation which is numerically solved becomes
\begin{multline}\label{eq:f12NumDifEq}
\left(\Bigl(\leftdu_x - \frac{\mu(x)}{2}\Bigr)\cdot\Bigl(\rightdu_x - \frac{\mu(x)}{2}\Bigr) + \nu(x)\right)\sqrt{w_{\phi}}f(x) \\
= E\,\sqrt{w_{\phi}}f(x).
\end{multline}
The solution on the grid is not reliable for the outer region, which is accounted for by solving the differential equation for the correlation function in the asymptotic region analytically (Ap.~\ref{ap:fAsymp}). The exponent  and the fractional power of the correlation function are directly related to the ones of the orbital and the total energy as
\begin{subequations}\label{eq:asympConnection}
\begin{align}\label{eq:gammaConnection}
\gamma_f &= \sqrt{-E} - \gamma_{\phi}, \\*
\label{eq:powerConnection}
p_f &= -\frac{(4p_{\phi}+1)\bigl(p_{\phi}\sqrt{-E} + 2\lambda_v\bigr) + p_{\phi}\lambda_w}{2p_{\phi}\sqrt{-E}}.
\end{align}
\end{subequations}
This analytical asymptotic solution is then glued smoothly to the numerical one by using the coefficients of the polynomial ($q_0$, $q_1$ and $q_2$ in~\eqref{eq:asympBehavior}) to fit the asymptotic solution to last three points of the inner region.
The last step is the full optimization. The combination of the HF orbital with its corresponding correlation function calculated in the second step is used as the starting guess. At short distances this is quite a good guess. However, for a stretched H--H bond the fully optimized orbital differs significantly from the HF orbital, so this starting guess is quite bad for this case. For elongated bonds, the converged result from a calculation with a similar bond length provides usually a better starting point.

The asymptotic behavior of the correlation function from a straightforward energy optimization does in general not agree with the relations in~\eqref{eq:asympConnection}. To cope with this deficiency we have added a penalty function to the energy, which is simply the disagreement in the exponent, $\Delta\gamma_f$, and power, $\Delta p_f$. So instead of the energy, we optimize the Lagrangian
\begin{align*}
L = E + c_{\text{pen}}\bigl((\Delta\gamma_f)^2 + (\Delta p_f)^2\bigr),
\end{align*}
where $c_{\text{pen}}$ controls how much we penalize for the mismatch. For good convergence one generally starts with a low value of $c_{\text{pen}}$ (0.0 for example) and lets the calculation converge. If the errors $\Delta\gamma_f$ and $\Delta p_f$ are too large, the value of $c_{\text{pen}}$ is increased and the calculation is restarted. A final value of $c_{\text{pen}} = 1.0\cdot10^3$ turned out to be sufficient for our purposes.

\section{Asymptotic behavior of $\mu(x)$ and $\nu(x)$}
\label{ap:asympMuAndNu}
To derive the asymptotic behavior of the function $\mu(x)$ and $\nu(x)$ appearing in the differential equation for the correlation function~\eqref{eq:fStat}, we use that the orbital behaves asymptotically as $x^p\e^{-\gamma x}$. In this appendix we drop the subscript $\phi$ for brevity. Since the correlation function does not occur, this does not lead to any confusion. To be able to evaluate the integrals, we have to resort to the easiest situation, $\rho=0$, so effectively only one atom in the system
\begin{align*}
\phi(x)	&= \abs{x}^p\e^{-\gamma\abs{x}}, \\
\phi'(x)	&= \sgn(x)\bigl(p - \gamma\abs{x}\bigr)\abs{x}^{p-1}\e^{-\gamma\abs{x}}, \\
\phi''(x)	&= \bigl(\gamma^2\abs{x}^2 - 2p\gamma\abs{x} + p(p-1)\bigr)\abs{x}^{p-2}
\e^{-\gamma\abs{x}} \notag \\
&\eqspace +
2\delta(x)\bigl(\delta_{p,1} - \gamma\delta_{p,0}\bigr)
\end{align*}
Further, the soft Coulomb potential is in general too hard to integrate analytically. Therefore, we use as a simplified version $v(x) = 2\lambda_v/\abs{x}$, which is identical to the soft-Coulomb potential in the asymptotic limit. Using these explicit expressions for the orbital, its derivatives and the potential, we can work out the required integrals as
\begin{widetext}
\begin{align*}
\binteg{y}{-\infty}{\infty}\phi^2(y)\phi^2(y+x)
&= \frac{\sqrt{\pi}\Gamma(2p+1)}{2^{4p+1}}\biggl(
\frac{x^{4p+1}\e^{-2\gamma x}}{\Gamma\bigl(2p + \frac{3}{2}\bigr)} + 
\frac{2}{\pi}\Bigl(\frac{x}{\gamma}\Bigr)^{2p+\half}\BesselK_{2p+\half}(2\gamma x)\biggr) \\
\binteg{y}{-\infty}{\infty}\phi^2(y)\phi(y+x)\phi'(y+x)
&= \frac{\sqrt{\pi}\Gamma(2p+1)}{2^{4p+2}}\biggl(
\frac{(1+4p-2\gamma x)x^{4p}\e^{-2\gamma x}}{\Gamma\bigl(2p+\frac{3}{2}\bigr)} - 
\frac{4\gamma}{\pi}\Bigl(\frac{x}{\gamma}\Bigr)^{2p+\half}\BesselK_{2p-\half}(2\gamma x)\biggr) \\
\binteg{y}{-\infty}{\infty}\phi^2(y)\phi(y+x)\phi''(y+x)
&= -2\gamma\delta_{p,0}\e^{-2\gamma x} + \frac{p\sqrt{\pi}\Gamma(2p-1)}{2^{4p}} \notag \\*
&\eqspace \times
\Biggl( 
\frac{\bigl(2p(p-1)(4p+1) + (1 + 2 p - 8 p^2) \gamma x  + (2p-1) \gamma^2x^2\bigr)
x^{4p-1}\e^{-2\gamma x}}{\Gamma\bigl(2p+\frac{3}{2}\bigr)} \notag \\*
&\eqspace\hphantom{{}\times\Biggl(} -
\frac{2}{\pi}\Bigl(\frac{x}{\gamma}\Bigr)^{2p-\half}\bigl(4p(p-1)\BesselK_{2p-\half}(2\gamma x) -
(2p-3)\gamma x\BesselK_{2p+\half}(2\gamma x)\Biggr) \\
\binteg{y}{-\infty}{\infty}\phi^2(y)\phi^2(y+x)v(y)
&= 2\lambda_v\frac{\sqrt{\pi}\Gamma(2p)}{2^{4p}}
\biggl( \frac{\e^{-2\gamma x} x^{4 p}}{\Gamma\bigl(2p+\half\bigr)} + 
\frac{2\gamma}{\pi} \Bigl(\frac{x}{\gamma}\Bigr)^{2p+\half} \BesselK_{2p+\half}(2\gamma x)\biggr),
\end{align*}
\end{widetext}
where $\Gamma(x)$ is the usual gamma function and $\BesselK_{\alpha}(x)$ denotes the modified Bessel function of the second kind. These integrals do not converge for arbitrary parameters. Of course we need $\gamma > 0$, but there are also constraints on the power $p$. In particular for the first integral we need $p > -\half$, for the second $p \geq 0$, for the third $p > \half$ or $p = 0$ and for the fourth $p > 0$. The last integral does not converge for $p=0$ due to the singularity of the potential $v(x)$. One can still obtain an explicit expression for $p=0$ using the potential $v(x) = 2\lambda_v/(\abs{x} + \abs{\alpha})$, though it is not so useful, since the third integral does not converge for $0 < p \leq \half$ and we want to vary $p$ in a smooth manner. Therefore, in general we require $p > \half$. The other required integrals can be obtained from the ones given before by noting that
\begin{multline*}
\integ{y}\phi^2(y)\phi(y-x)\phi'(y-x) \\
= \integ{y}\phi^2(y)\phi(y+x)\phi'(y+x)
\end{multline*}
and
\begin{multline*}
\integ{y}\phi^2(x+y)\phi(y)\phi''(y) \\
= \integ{y}\phi^2(y)\phi(y+x)\phi''(y+x).
\end{multline*}
Using that the modified Bessel function of the second kind for $x \gg \abs{\alpha^2-1/4}$ behave as
\begin{align}
\BesselK_{\alpha}(x) \approx \sqrt{\frac{\pi}{2x}}\e^{-x}\biggl(1 + \frac{4\alpha^2-1}{8x} + 
\Order\bigl(1/x^2\bigr)\biggr),
\end{align}
we can work out the asymptotic behavior of $\mu(x)$ for $x \to \infty$ as
\begin{multline}\label{eq:muAsymp}
\mu(x) = -2\gamma + \frac{4p + 1}{x} \\*
 - \frac{(2p+1)\Gamma\bigl(2p + \frac{3}{2}\bigr)}
{\gamma^{2p+1}\sqrt{\pi}}\frac{1}{x^{2p+2}} + \Order\bigl(1/x^{2p+3}\bigr).
\end{multline}
The function $\nu(x)$ has components from the kinetic energy, potential energy and the interaction.  The asymptotic behavior of the kinetic contribution can be worked out to be
\begin{subequations}\label{eq:nuAsymp}
\begin{align}
\nu_t(x)
&= -\gamma^2 - (4p+1)\gamma\frac{1}{x} - \frac{2p(p-1)(4p+1)}{2p-1}\frac{1}{x^2} \notag \\*
&\eqspace +
\frac{\Gamma\bigl(2p+\tfrac{3}{2}\bigr)}{\gamma^{2p}\sqrt{\pi}}\biggl(
\frac{2}{2p-1}\frac{\gamma}{x^{2p+1}} - \frac{p+1}{x^{2p+2}}\biggr) \notag \\*
&\eqspace + 
\Order\bigl(1/x^{2p+3}\bigr)
\end{align}
and likewise for the potential contribution we have
\begin{align}
\nu_v(x)
&= 2\lambda_v\frac{4p+1}{p}\frac{1}{x} + 
2\lambda_v\frac{\Gamma\bigl(2p+\frac{3}{2}\bigr)}{\gamma^{2p}\sqrt{\pi}}\notag \\*
&\eqspace \times
\biggl(\frac{2}{p}\frac{1}{x^{2p+1}} +
\frac{2p^2 - 3p - 1}{p\gamma}\frac{1}{x^{2p+2}}\biggr) \notag \\*
&\eqspace +
\Order\bigl(1/x^{2p+3}\bigr).
\end{align}
\end{subequations}

\section{Asymptotic solution of the correlation function}
\label{ap:fAsymp}
Since the asymptotic part of the correlation is hard to solve numerically, it should be calculated analytically. The first step is to note that the function $\mu(x)$ and $\nu(x)$ become constant for large $x$ (Eqs~\eqref{eq:muAsymp} and~\eqref{eq:nuAsymp}), so the differential equation for the correlation function~\eqref{eq:fStat}, reduces in the far asymptotic region to
\begin{align*}
\bigl(\du_x^2 - 2\gamma_{\phi}\du_x - \gamma_{\phi}^2\bigr)f(x) = E\,f(x).
\end{align*}
This differential equation is readily solved by $f(x) = \e^{-\gamma_f x}$, with $\gamma_f = -\gamma_{\phi} \pm \sqrt{-E}$. However, for the non-interacting case we know that $\gamma_f = 0$ and that $\gamma_{\phi}^2 = -E$, so we need to choose the plus-sign. Hence we find that $\gamma_f = \sqrt{-E} - \gamma_{\phi}$~\eqref{eq:gammaConnection}. To refine our asymptotic solution, we need to include higher order terms. Including the asymptotic behavior of the functions $\mu(x)$ and $\nu(x)$ up to first order in $1/x$ gives the following asymptotic equation
\begin{align*}
\biggl(-\du_x^2 + \Bigl(2\gamma_{\phi} - \frac{4p_{\phi}+1}{x}\Bigr)\du_x
- \gamma_{\phi}^2 + \frac{\bar{\lambda}}{x}\biggr)f(x) = E\,f(x),
\end{align*}
where we have introduced
\begin{align*}
\bar{\lambda} \coloneqq (4p_{\phi}+1)\gamma_{\phi} + \frac{4p_{\phi}+1}{p_{\phi}}2\lambda_v + \lambda_w.
\end{align*}
To use our previous asymptotic solution, we write the correlation function as $f(x) = \tilde{q}(x)\e^{-\gamma_fx}$. Using this ansatz, we obtain the following differential equation for $\tilde{q}(x)$
\begin{align*}
\biggl(-\du_x^2 + \Bigl(2\sqrt{-E} - \frac{4p_{\phi}+1}{x}\Bigr) + 
\frac{\tilde{\lambda}}{x}\biggr)\tilde{q}(x) = 0,
\end{align*}
where
\begin{align*}
\tilde{\lambda} &\coloneqq \bar{\lambda} + (4p_{\phi}+1)\gamma_f \notag \\*
&\hphantom{:}= (4p_{\phi}+1)\sqrt{-E} + \frac{4p_{\phi}+1}{p_{\phi}}2\lambda_v + \lambda_w.
\end{align*}
Due to the $\bar{\lambda}/x$-term a straightforward expansion in $1/x$ will not work, since no term from the expansion can cancel the $1/x$ behavior. Therefore, we use the Frobenius trick: we write $\tilde{q}(x) = x^{p_f}q(x)$ and choose $p_f$ such that it cancels the $\bar{\lambda}/x$-term. Working out the equations, one finds that $p_f = -\tilde{\lambda}/(2\sqrt{-E})$, which can be worked out further to give~\eqref{eq:powerConnection}. Although we do not use the explicit solution for $q(x)$ in the calculations, we still include it here for completeness. The remaining differential equation for $q(x)$ becomes
\begin{multline*}
\biggl(-\du_x^2 + \Bigl(2\sqrt{-E} - \frac{4p_{\phi}+2p_f+1}{x}\Bigr)\du_x \\
- \frac{p_f(4p_{\phi}+p_f)}{x^2}\biggr)q(x) = 0.
\end{multline*}
The function $q(x)$ is expressed as a power-series in $1/x$
\begin{align*}
q(x) = \sum_{k=0}^{\infty}\frac{q_k}{x^k}.
\end{align*}
Inserting the power series in the differential equation, we obtain a recursion relation between the consecutive coefficients
\begin{align*}
q_{k+1} = \frac{(4p_{\phi}+p_f-k)(k-p_f)}{2\sqrt{-E}(k+1)}q_k
\end{align*}
and $q_0 \neq 0$ determines the overall scaling of the correlation function via $q(x)$.

\bibliography{bible}

\end{document}